\documentclass[journal,article,submit,moreauthors,pdftex]{Definitions/mdpi} 
\usepackage{amssymb,amsfonts,textcomp}
\usepackage{array}
\usepackage{supertabular}
\usepackage{hhline}
\usepackage{graphicx}
\preto{\abstractkeywords}{\nolinenumbers}
\newcommand{\av}[1]{\langle {#1} \rangle}

\Title{Power-law distributions of dynamic cascade failures in power-grid models}

\Author{G\'eza \'Odor,  B\'alint Hartmann }
\AuthorNames{G\'eza \'Odor, B\'alint Hartmann}

\address{
Centre for Energy Research, P. O. Box 49, H-1525 Budapest, 
Hungary; odor.geza@energia.mta.hu}

\corres{Correspondence: odor.geza@energia.mta.hu}

\abstract{
Power-law distributed cascade failures are well known in power-grid systems.
Understanding this phenomena has been done by various DC threshold models,
self-tuned at their critical point. Here we attempt to describe it using an
AC threshold model, with a second-order Kuramoto type equation of motion
of the power-flow.
We have focused on the exploration of network heterogeneity effects,
starting from homogeneous 2D square lattices to the US power-grid, possessing
identical nodes and links, to a realistic electric power-grid obtained
from the Hungarian electrical database. The last one exhibits node dependent
parameters, topologically marginally on the verge of robust networks.
We show that too weak quenched heterogeneity, coming solely from the
probabilistic self-frequencies of nodes (2D square lattice) is not sufficient
to find power-law distributed cascades. On the other hand too strong 
heterogeneity destroys the synchronization of the system.
We found agreement with the empirically observed power-law failure size
distributions on the US grid, as well as on the Hungarian networks near
the synchronization transition point. We have also investigated the
consequence of replacing the usual Gaussian self-frequencies to exponential
distributed ones, describing renewable energy sources. We found a drop
in the steady state synchronization averages, but the cascade size distribution 
both for the US and Hungarian systems remained insensitive and have kept the 
universal tails, characterized by the exponent $\tau\simeq 1.8$. 
We have also investigated the effect of an instantaneous feedback mechanism in
case of the Hungarian power-grid.
}


\begin{document}

%%%%%%%%%%%%%%%%%%%%%%%%%%%%%%%%%%%%%%%%%%%%%%%%%%%%%%%%%%%%%%%%%%%%%%%%%
\section{Introduction}
%%%%%%%%%%%%%%%%%%%%%%%%%%%%%%%%%%%%%%%%%%%%%%%%%%%%%%%%%%%%%%%%%%%%%%%%%

Modeling power grids has become a hot topic in statistical physics
as electric energy infrastructure is bound to undergo huge changes in
both the generation and demand sides to make it environmentally
sustainable. They are large complex, heterogeneous dynamical system,
built up from nodes of energy suppliers and consumers, interconnected
by a network with hierarchical modular (HMN) structure~\cite{Acebron,ARENAS200893,POWcikk}.
The transition from fossil to renewable energy sources poses unprecedented
challenges towards the robustness and resilience of power grids
as they introduce correlated spatio-temporal fluctuations.

Unexpected changes may cause desynchronization cascades, propagating through 
the whole system as an avalanche, causing blackouts of various sizes.
These can lead to full system desynchronization lasting for long time 
\cite{1}. 
Numerous attempts have been made for understanding and forecasting power 
outages from several different angles~\cite{Rev-Eng}.
Particularly, from the point of view of statistical physics of breakdown 
phenomena, systemic risk of failure in power infrastructure 
represents a particular case of a generic phenomena: the risk of system-wide 
breakdown in threshold activated disordered systems.

The size distributions of the outages have been found scale-free in the US, 
China, Norway, Sweden in the available long time series data~\cite{Car}. 
They have been modeled~\cite{Car2} by direct current (DC) threshold models
with self-organized criticality (SOC)~\cite{SOC}, arising as the consequence 
of self-tuning to a critical point by the competition of power demand and 
network capabilities. These models are similar to those of sand piles, in which 
redistribution avalanches are generated, when the local level exceeds a 
threshold value.
By analyzing the statistics of 7 years (from 2002 to 2008) of EU network 
failures a moderate support for scale-free behavior has
been found~\cite{ROSASCASALS2011805}. In particular, power-laws could
be fitted better in countries, with so called robust networks~\cite{CasTh}.
The categorization of robust/fragile is based on the static network 
topology analysis of national power-grids~\cite{Pgridtop},
where networks with $P(k > K) = C \exp(-k/\gamma)$ cumulative degree ($k$)
distribution and $\gamma < 3/2$ are called robust.
Restoration time was supported particularly well by a power law (PL) model
in both groups, but this behavior is in accordance with findings, 
where human temporal response distributions have been found to be fat 
tail distributed. It is well known that human behavior exhibits
bursty behavior~\cite{burstcikk}, which raises the question whether
the observed PL-s are the consequence of the power-grid function itself or
related to the bursty behavior of system maintenance procedures. 
One of the aims of our study is to investigate if such PL-s can be 
reproduced by more realistic power-grid models than the first 
attempts made using simple threshold ones.
 
The framework of direct current DC threshold models~\cite{Car2} can be 
extended by taking into account the real power flow in alternating current
(AC) networks by modeling via the second order Kuramoto equation ~\cite{fila}. 
A number of studies exists, which focus on the synchronization and stability 
issues. 
In~\cite{Carareto} the authors show that the coupling strength of power grid 
models behaves differently, depending on the heterogeneity of 
the nodes; synchronization appears in highly heterogeneous complex networks, 
where nodes show different characteristics.
Our work considers even more heterogeneous system derived from real data 
and goes beyond the bimodal Gaussian self-frequency approximations too.
Choi et al.~\cite{CHOI201132,5} 
use various frameworks to test the effect of inertia on the speed of 
synchronization. Their results imply that large inertia induces slower 
synchronization. In their works, D\"orfler et al.~\cite{12,13} examine 
synchronization and stability in power networks and other complex networks, 
applying non-uniform (heterogeneous) parameters to the Kuramoto-model. 
Using the real topology of the Italian transmission network, 
Fortuna et al.~\cite{18} find that the class of Kuramoto-like models 
with bimodal distribution (sources and consumers) of the frequencies 
is the most appropriate mapping between oscillators and power system nodes. 
The same network and modeling approach is also used in~\cite{29}, 
concluding that the synchronization transition is hysteric for 
sufficiently large masses, but for Italian high voltage  power grid
the transition is largely nonstrategic, due to the low value of the 
average connectivity.  
Future spread of distributed generation is modeled in~\cite{32}, using 
non-uniform parameters for power system nodes. Results of the authors 
show that realistic (non-optimal) topologies have wider phase differences 
between connected nodes, which leads to less homogeneously transmitted 
power, but no significant differences have been observed in case of 
node removal. 
Smaller topologies are used in~\cite{36} to test the extension of the 
Kuramoto-model with voltage dynamics to study the voltage-angle 
stability of power systems. The authors of~\cite{Gryz} introduce a 
method to estimate coupling strength of power grids, which is a 
crucial parameter of the Kuramoto-model; proper knowledge of such 
parameters can help maintaining stability of the power system even in 
the presence of large transients. 
The recent work by Taher et al.~\cite{Olmi-Sch-19} proposes a time-delayed 
feedback control to the Kuramoto-model and test it on a realistic topology, 
with complex bimodal self-frequency distributions.

Our study goes beyond the synchronization stability issues, by generating
failure cascade distributions, which had been considered in DC models only.
Solving AC power flow equations is a significant computational challenge. 
The DC approach limits this by linearizing the equations and has been used 
in large-scale simulations. It considers active powers, but ignores 
reactive ones and transmission losses~\cite{6819069}. Its efficiency 
approximates the AC power flow, without being iterative and complex
\cite{doi:10.1063/1.4807478,doi:10.1111/risa.12281}. 
It misrepresents transmission line flows by less than $5\%$, but about 
10 times faster than the exact solution provided by the AC load flow 
approach~\cite{ACflowmod}. 
However, while for DC threshold models SOC critical transition is established, 
for AC threshold models we have no knowledge how the underlying second order 
Kuramoto model, which has a first order transition~\cite{POWcikk}, 
affects the avalanche size distributions. One of the main objective of our
study is to show how scale-free avalanches can occur in the AC Kuramoto
threshold model as we increase the network heterogeneity.

The synchronization and stability can be deduced from the power transfer 
behavior of a load/supply AC electrical circuit and turns out to be the 
generalization of the Kuramoto model~\cite{kura} with inertia. 
The Kuramoto model below $d < d_l=4$ does not exhibit real phase transition 
to a synchronized state, but a smooth crossover only~\cite{chate_prl}. 
In real life we can observe partially synchronized states. 
The second order Kuramoto equation is also expected to have $d_l=4$,
and in lower graph dimensions the transition point shifts to infinity
with the system size and hysteresis behavior emerges~\cite{POWcikk}.

While most of the SOC models are homogeneous, which means all nodes 
and interactions are the same and the connection matrix is regular 
deterministic, in real life all kinds of heterogeneity can occur in 
the connection network topology as well as in the node/interaction parameters. 
Highly heterogeneous, also called disordered with respect to the 
homogeneous, system can experience rare-region effects altering
critical dynamics~\cite{Vojta}. These rare regions, which are locally 
in another state than the whole, evolve slowly and contribute to 
the global order parameter, causing slow dynamics and fluctuations.
They can generate so-called Griffihts Phases (GP)~\cite{Griffiths} 
in an extended region around the critical point, causing slowly
decaying auto-correlations and burstyness \cite{burstcikk}.
In synchronization models such rare regions can cause frustrated
synchronization and chimera states~\cite{Frus,Frus-noise,FrusB}.
These result in non-universal PL distributions of the desynchronization 
events below the transition point~\cite{POWcikk,KurCC,KKI18deco}.
In Ref.~\cite{POWcikk} we provided numerical evidence for this by 
modeling a sudden drop of global coupling of the second order Kuramoto
model defined on 2D square lattices and on large synthetic power-grids.

Very recently dynamical modeling of cascade failure has been 
introduced combining the second order Kuramoto with power transfer 
thresholds~\cite{SWTL18}. Identification of critical lines of 
transmission in different national power grids has been determined.
We follow this method in order to investigate the desynchronization
duration distributions via measuring the number of failed lines 
following a node removal event. 
We shall compare results obtained on 2D square lattices with those of the
US high voltage power-grid and the Hungarian power-grid with 418 nodes 
that we generated from our network providers.

Modeling power-spectra of renewable energy sources has been done
in case of wind farms and solar cells~\cite{Anvari_2016}. The effects
of sudden weather changes and the strong spatio-temporal correlations
decrease the stability of power grids. The power output of a single
unit deviates largely from the normal distribution, but this 
non-Gaussian behavior remains also for the aggregated power of farms. 
Therefore, the central limit theorem, predicting a convergence to 
Gaussian for independent data sets with defined standard
deviation, does not apply. We shall also investigate here the effects of
replacing Gaussian self-frequency distributions to exponential ones
in case of our power-grid models. In particular, we test the robustness
of the scale-free behavior of outage distributions, by the replacement
of all nodes to non-Gaussian.

%%%%%%%%%%%%%%%%%%%%%%%%%%%%%%%%%%%%%%%%%%%%%%%%%%%%%%%%%%%%%%%%%%%%%%%%%
\section{Models and methods}
%%%%%%%%%%%%%%%%%%%%%%%%%%%%%%%%%%%%%%%%%%%%%%%%%%%%%%%%%%%%%%%%%%%%%%%%%

The main purpose of using the Kuramoto-model is to examine cascade failures. 
Transmission System Operators traditionally use a static approach for such 
analysis, which means that they start the simulation at a fixed operating 
point by performing a load-flow, trip the faulty line (remove the edge 
from the graph) and then perform another load-flow at this different operating 
point. While this method is simple, it fails to capture the dynamic response 
of the units (generators and loads) in the system, since the iterative nature 
of load-flow calculations aims to create a numerical solution; if necessary, 
by linearization and simplification. 
In the contrary, the Kuramoto-model starts the simulation at a fixed operating 
point by performing thermalization (which is a dynamic process), trip the 
faulty line and examine the unfolding transient, which will reveal dynamic 
response of the units.

The evolution of synchronization is based on the swing equations~\cite{swing}
set up for mechanical elements with inertia by the second order 
Kuramoto equation~\cite{fila}.
For a network of $N$ oscillators with phase $\theta_i(t)$:
\begin{eqnarray}\label{kur2eq}
\dot{\theta_i}(t) & = & \omega_i(t) \\
\dot{\omega_i}(t) & = & \omega_i^0 - \alpha \dot{\theta_i}(t) 
+ K \sum_{j=1}^{N} A_{ij} \sin[ \theta_j(t)- \theta_i(t)] \ , \nonumber
\end{eqnarray}
where $\alpha$ is the damping parameter, describing the power dissipation,
$K$ is the global coupling, related to the maximum transmitted power
between nodes and $A_{ij}$, which is the weighted adjacency matrix
of the network, containing admittance elements.
Very recently this equation has been refined with the aim of application 
for the German HV power-grid by~\cite{Olmi-Sch-19} 
\begin{equation}\label{1}
{\ddot{{\theta }}}_{i}+\alpha {\ }{\dot{{\theta
}}}_{i}=\frac{{P}_{i}}{{I}_{i}{\ }{\omega
}_{G}}+\frac{{K}_{i}}{{I}_{i}{\ }{\omega }_{G}}{\ }\sum
_{j=1}^{N}{{A}_{\mathit{ij}}{\ }\sin \left({\theta }_{j}-{\theta
}_{i}\right)} \ .
\end{equation}
Generator units (${P}_{i}>0$) and loads ( ${P}_{i}<0$) are modeled with 
a bi-modal probability distribution with peaks at mean values of power
sources and sink.  
The authors assume homogeneous transmission capacities, thus  ${K}_{i}=K$. 
The dissipation parameter $0.1 \ [1/s] \le \alpha \le 1 \ [1/s]$ and moments of 
inertia at the nodes is also considered to be homogeneous: 
${I}_{i} = I = 40{\ }{10}^{3} \ [\mathit{kg}{m}^{2}]$, 
which approximately equals the moment of inertia of a $400$ MW power plant. 
The adjacency matrix is constructed of binary elements, $1$ represents 
connection, $0$ represents the lack of it.
The authors cite that previous applications of the Kuramoto equation had
a significant limitation as all generators and loads were handled with
a bi-modal $\delta ${}-distribution, where all units had the same
power. However, the proposed method uses empirical data for
${P}_{i}$ only and all other parameters are handled in a uniform way. 
In the following we extend this as follows.

Considering Eq.~(\ref{1}), the following statements can be made:
\begin{enumerate}
\item  $\alpha $ dissipation factor is chosen to be equal to $0.4 / [1/s]$,
which value will be used in this paper as well
\item in real power systems, the $i$-th node has connection both to
generators and loads, thus  ${P}_{i}$ parameter of the equation can be
written as
\end{enumerate}
\begin{equation}\label{2}
{P}_{i}={P}_{\mathit{Gi}}-{P}_{\mathit{Li}} \ ,
\end{equation}
where ${P}_{G}$ represents generators (production),  ${P}_{L}$
represents loads (consumption). 

For a given node, the ratio of  ${P}_{Gi}$ and  ${P}_{Li}$ shows
significant dependence on the voltage of the node and the size of the
supplied service area. If the node serves as the connection point of a
power plant,  ${P}_{Li}{\cong}0$, since only self-consumption of the
plants has to be considered as a load. If the node only supplied
consumers,  ${P}_{Gi}=0$ \footnote{It has to be noted that due to the
increasing number of distributed generators, such purely consuming
nodes are becoming less frequent.}. The third case is the most typical,
when the node connects both supplies and loads. In such cases the
ratio of ${P}_{\mathit{Gi}}$ and  ${P}_{\mathit{Li}}$ will determine
not only that a certain node will behave as a net producer or a net
consumer, but also the moment of inertia for that service area. Exact
ratios might also depend on actual load state, season, day of the week,
etc., which variations could be addressed by using so-called
characteristic load states (summer and winter peak e.g.).

The ${I}_{i}$ moment of inertia can be considered as a sum of two
contributions: inertia of generators and inertia of loads. In large
power systems the cumulative moment of inertia of power plants exceeds
that of the loads by magnitudes, so load inertia is often neglected. In
the examined network model however there are numerous subsystems, where
the power (and thus the inertia) of generators is very low or even
zero. The relation between the body moment of inertia, apparent power
$S_i$ and $H$ inertia constant is: 
\begin{equation}\label{3}
{I}_{i}=\frac{2{\ }H{\ }{S}_{i}}{{{\omega }_{G}}^{2}}
\end{equation}
The magnitude of the inertia constant is highly dependent on the type of the
power plant (see Table~\ref{tab:InertiaG}) and the load mix 
(see Table~\ref{tab:InertiaC}) as well, thus uniform handling of ${I}_{i}$ 
is a simplification of modeling.

\begin{table}
 \caption{\label{tab:InertiaG} Typical inertia constant of power plant types}
 \centering
 \begin{tabular}{|c|c|}
  \hline
  Production type  & $H[s]$ \\
  \hline
   Nuclear              & $6$ \\
   Combined cycle gas turbine & $5.5$ \\
   Single-shaft gas turbine & $4.5$ \\
   Large-scale hydro &  3 \\
   Diesel genset & 2 \\
   Converter-based units & 0 \\
 \hline
 \end{tabular}
\end{table}

\begin{table}
 \caption{\label{tab:InertiaC} Typical inertia constant of certain consumers~\cite{Cons-H}}
 \centering
 \begin{tabular}{|c|c|}
  \hline
  System  & $H[s]$ \\
  \hline
  Direct-on-line induction motor and compressor & 1\\
  Direct-on-line induction motor and conveyor belt & 0.6\\
  Direct-on-line synchronous motor and compressor & 1\\
  Variable speed drive & 0 \\
  Lighting & 0\\
\hline
 \end{tabular}
\end{table}

The ${K}_{i}$ couplings represent the amount of power that can be
transmitted from the $i$-th node. If elements of ${A}_{\mathit{ij}}$
adjacency matrix take up binary ($0/1$) values, the dimension of the
coupling is power:  $\left[{K}_{i}\right]=\mathit{MW}$. Such power
values are usually available in the database of system operators as
operation limits. These operational limits can be based on thermal
limits (to avoid overloading of the conductor) or limited capabilities
of the infrastructure (measurement transformers, switch gear, etc.).
Operational limits show large dependence on voltage level, age of the
infrastructure and seasons, thus uniform handling of this parameter
is also a simplification of modeling. 
In conclusion, returning to the equation by~\cite{Olmi-Sch-19} for 
${P}_{i}$, ${K}_{i}$ and ${I}_{i}$ empirical distribution values can 
be used instead of an uniform characterization.

Taking into consideration that multiple generators and loads can be
connected to the same node, cumulative values (e.g. net load) will be
marked by $\mathit{area}, i$ index instead of the $i$ index. 
Transforming Eq.~(\ref{1}), ${P}_{\mathit{area},i}$ will represent 
the net load of a certain area:
\begin{equation}\label{5}
{\ddot{{\theta }}}_{\mathit{area},i}+\alpha {\ }{\dot{{\theta
}}}_{\mathit{area},i}=\frac{{P}_{\mathit{area},\mathit{Gi}}-{P}_{\mathit{area},\mathit{Li}}}{{I}_{\mathit{area},i}{\ }{\omega
}_{G}}+\frac{{K}_{\mathit{area},i}}{{I}_{\mathit{area},i}{\ }{\omega
}_{G}}{\ }\sum _{j=1}^{N}{{A}_{\mathit{ij}}{\ }\sin
\left({\theta }_{j}-{\theta }_{i}\right)}
\end{equation}
Using the relation~\ref{3}, we are able to express the inertia constant of
the service area:
\begin{equation}\label{6}
{I}_{\mathit{area},i}=\frac{2{\ }{H}_{\mathit{area}}{\ }{S}_{\mathit{area},i}}
{{{\omega}_{G}}^{2}} \ ,
\end{equation}
If the area only consists of generators, ${H}_{\mathit{area}}={H}_{\mathit{Gi}}$ 
and the value can be determined based on the composition of the power 
plant portfolio, using Table~\ref{tab:InertiaG}. For European
power systems, these values are expected to be between $6$ s and $1$ s in
2030, depending on the power plant portfolio~\cite{ENTSO}. 
If the area consists of both generators and loads, the value of  
${I}_{\mathit{area},i}$ can be calculated taking into consideration 
inertial response of both generators and loads
\begin{equation}\label{7}
{I}_{\mathit{area},i}=\frac{\sum
{\left(2{\ }{H}_{\mathit{Gi}}{\ }{S}_{\mathit{Gi}}\right)}}{{{\omega
}_{G}}^{2}}+\frac{\sum
{\left(2{\ }{H}_{\mathit{Li}}{\ }{S}_{\mathit{Li}}\right)}}{{{\omega
}_{G}}^{2}} \ ,
\end{equation}
where  ${S}_{\mathit{Gi}}$ is the power of single generator units,
${S}_{\mathit{Li}}$ is the power of single load units. Value of
${H}_{\mathit{Gi}}$ can be chosen from Table~\ref{tab:InertiaG}, 
while in case of ${H}_{\mathit{Li}}$ certain empirical values 
can be used (see Table~\ref{tab:InertiaC}). 
In this paper it is assumed, that 60-70\% of total load is of
rotating machines ($H = 0.5$ \ [s]), and the remaining 30-40\% load is of
low inertia units ($H = 0.1$ \ [s]),  ${H}_{\mathit{Li}}$ equals:
\begin{eqnarray}
{H}_{\mathit{Li}} &=& \frac{\sum
{\left(0.5 \ s{\ }{S}_{\mathit{Li}}{\ }\left[0.6{\dots}0.7\right] + 0.1 \ s {\ }{S}_{\mathit{Li}}{\ }
\left[0.4{\dots}0.3\right]\right)}}{{S}_{\mathit{Li}}} \\ & = & \sum
{0.5 \ s{\ }\left[0.6{\dots}0.7\right]+0.1 {\ } s\ \left[0.4{\dots}0.3\right]}=\left[0.34{\dots}0.38\right] \ s {\approx} \ 0.36 {\ } s
\end{eqnarray}

\begin{figure}[H]
\centering
\includegraphics[width=10cm]{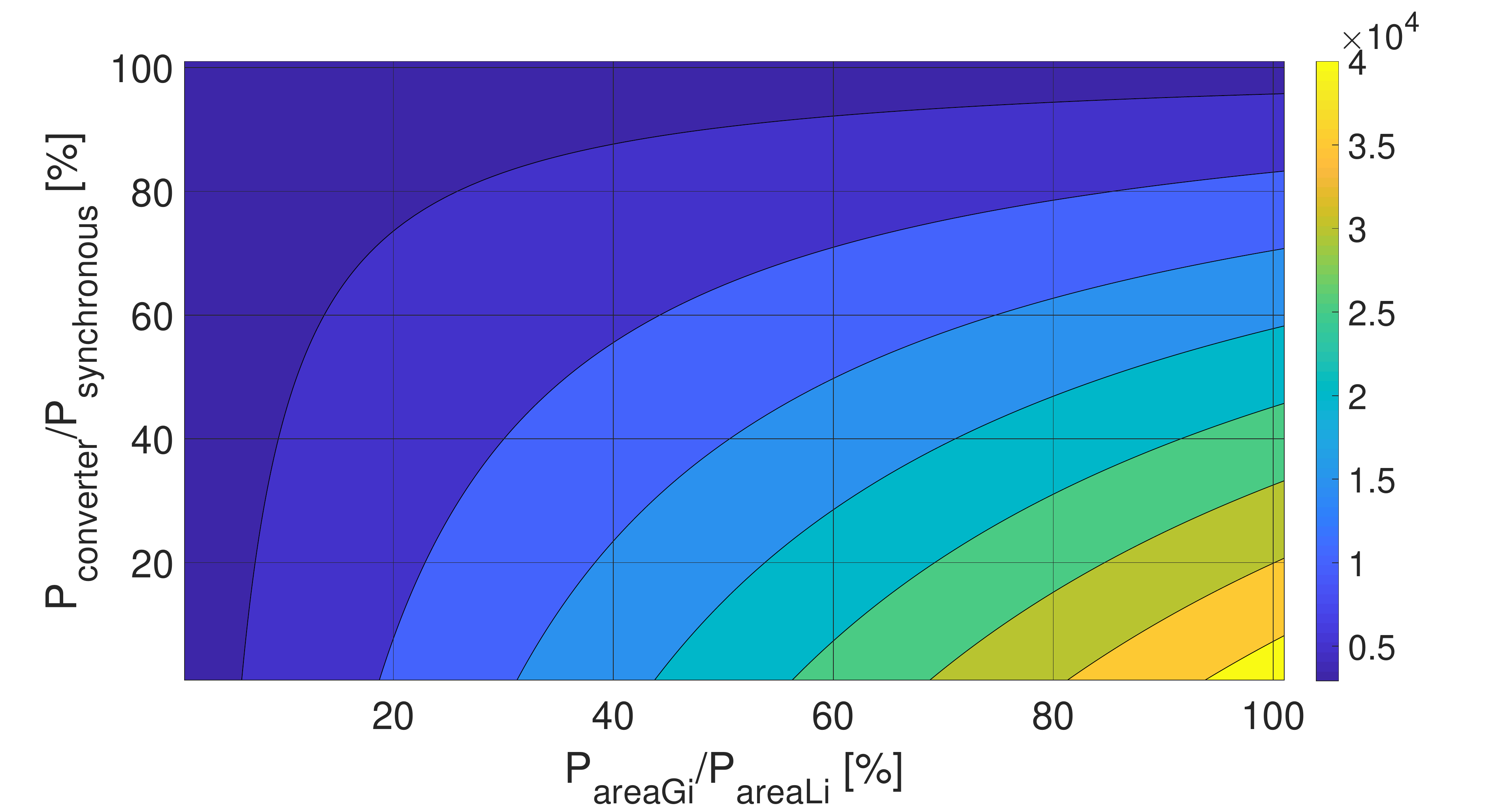}
\caption{Moment of inertia for a 400 MW node, depending on the proportion 
of locally generated power and the share of converter-based 
generation units.\label{X}}
\end{figure}

To underline the importance of properly assessing ${I}_{\mathit{area},i}$, 
an illustrative example is shown. In the paper by~\cite{Olmi-Sch-19},  
${I}_{i}=I=40 \times {10}^{3} \ [\mathit{kg}{m}^{2}]$ was used
as a representation of a $400$ MW power plant, which by substituting into
Eq.~(\ref{6}) will result  ${H}_{\mathit{area}}=4.93[s]$; this will be used
as  ${H}_{\mathit{Gi}}$ in the following example. Figure~\ref{X} shows how
the moment of inertia varies for a $400$ MW node, depending on the
proportion of locally generated power and the share of converter-based
generation units, which have no inertia. Values on the figure vary
between $2918$ and $42879 \ [\mathit{kg}{m}^{2}]$, which emphasizes the 
importance of using different inertia values for the nodes in such models. 
E.g. in case of the Hungarian model, only $10\%$ of the nodes can be 
represented as purely generation ones and the remaining $90 \%$ has 
substantially smaller moment of inertia.
%%%%%%%%%%%%%%%%%%%%%%%%%%%%%%%%%%%%%%%%%%%%%%%%%%%%%%%%%%%%%%%%%%%%%%%%%
%%%%%%%%%%%%%%%%%%%%%%%%%%%%%%%%%%%%%%%%%%%%%%%%%%%%%%%%%%%%%%%%%%%%%%%%

If we substitute Eq.~(\ref{7}) to the right side of Eq.~(\ref{5})
\begin{equation}\label{8}
\frac{{P}_{\mathit{area},\mathit{Gi}}-{P}_{\mathit{area},\mathit{Li}}}{{I}_{\mathit{area},i}{\ }{\omega
}_{G}}=\frac{{P}_{\mathit{area},\mathit{Gi}}-{P}_{\mathit{area},\mathit{Li}}}{\frac{\sum
{\left(2{\ }{H}_{\mathit{Gi}}{\ }{S}_{\mathit{Gi}}\right)}}{{{\omega
}_{G}}^{2}}+\frac{\sum
{\left(2{\ }{H}_{\mathit{Li}}{\ }{S}_{\mathit{Li}}\right)}}{{{\omega
}_{G}}^{2}}{\ }{\omega }_{G}}
\end{equation}
After simplification we get:
\begin{equation}\label{9}
\frac{\left({P}_{\mathit{area},\mathit{Gi}}-{P}_{\mathit{area},\mathit{Li}}\right){\ }{\omega
}_{G}}{2{\ }\sum
{\left({H}_{\mathit{Gi}}{\ }{S}_{\mathit{Gi}}\right)+\left({H}_{\mathit{Li}}{\ }{S}_{\mathit{Li}}\right)}}
\end{equation}
Assuming that the power factor is one ( $ S{\approx} P $):
\begin{equation}\label{10}
\frac{\left({P}_{\mathit{area},\mathit{Gi}}-{P}_{\mathit{area},\mathit{Li}}\right){\ }{\omega
}_{G}}{2{\ }\sum
{\left({H}_{\mathit{Gi}}{\ }{P}_{\mathit{Gi}}\right)+\left({H}_{\mathit{Li}}{\ }{P}_{\mathit{Li}}\right)}}=\frac{2{\ }\pi
{\ }50\mathit{Hz}}{2}{\ }\frac{\left({P}_{\mathit{area},\mathit{Gi}}-{P}_{\mathit{area},\mathit{Li}}\right)}{\sum
{\left({H}_{\mathit{Gi}}{\ }{P}_{\mathit{Gi}}\right)+\left({H}_{\mathit{Li}}{\ }{P}_{\mathit{Li}}\right)}}
\end{equation}
which shows that this part of Eq.~(\ref{7}) is affected by both generation and
load mix.

With similar steps, the remaining elements of Eq.~(\ref{5}) can be rewritten:
\begin{eqnarray}\label{11}
{\ddot{{\theta }}}_{\mathit{area},i}+\alpha {\ }{\dot{{\theta}}}_{\mathit{area},i} 
&=& \frac{{\omega}_{G}}{2}{\ }\frac{\left({P}_{\mathit{area},\mathit{Gi}}-{P}_{\mathit{area},\mathit{Li}}\right)}
{\sum{\left({H}_{\mathit{Gi}}{\ }{P}_{\mathit{Gi}}\right)+\left({H}_{\mathit{Li}}{\ }{P}_{\mathit{Li}}\right)}} \\ \nonumber
&+& \frac{{\omega}_{G}}{2}{\ }\frac{{K}_{\mathit{area},i}}
{\sum{\left({H}_{\mathit{Gi}}{\ }{P}_{\mathit{Gi}}\right)+\left({H}_{\mathit{Li}}{\ }{P}_{\mathit{Li}}\right)}}{\ }
\sum_{j=1}^{N}{{A}_{\mathit{ij}}{\ }\sin \left({\theta }_{j}-{\theta}_{i}\right)}
\end{eqnarray}
Eq.~(\ref{11}) is the form, which we used in the simulation code of 
Hungarian High Voltage (HU-HV) power-grid.

We have studied three different types of networks, by gradually increasing the
heterogeneity:
\begin{itemize}
\item $2D$ square lattices, with periodic boundary conditions, simulating 
homogeneous electric power-grids using Eq.~(\ref{kur2eq}).
\item The $4941$ node power-grid of the western states of the US
(US-HV)~\cite{USpg} with Eq.~(\ref{kur2eq}).
\item A $418$ node Hungarian HV electric power grid, deduced from
the Hungarian Transmission System Operator (MAVIR) detailed database, 
using Eq.~(\ref{11}).
\end{itemize}   

We evaluated at each time step the actual power flow along the 
transmission lines and compared it the available capacity of the edges of 
the network as in ~\cite{SWTL18}. The flow of the power from edge $j$ to 
$i$ with the generalized coupling
\begin{equation} 
K'_{ij} = \frac{\omega_{G}}{2}{\ }\frac{ K_{\mathit{area},i} A_{\mathit{ij} }} {\sum{\left({H}_{\mathit{Gi}}{\ }{P}_{\mathit{Gi}}\right)
+ \left( H_{\mathit{Li}}{\ } P_{\mathit{Li}}\right)}}
\end{equation}
is described by 
\begin{equation}
F_{ij} = K'_{ij} \sin \left( \theta_j - \theta_i \right) \ .
\end{equation}
The overload condition is expressed by a comparison with a fraction 
$T\in [0,1]$ of the maximum flow
\begin{equation}\label{thr-cond}
| F_{ij} | > T  K'_{ij} \ .
\end{equation}
During the solution of the equation of motion we checked this condition at
each time step. In case the power flow of the line exceeded a pre-set threshold,
we cut the line by resetting the adjacency matrix elements $A_{ij} = A_{ji}=0$.
These thresholds can be selected by the settings of transmission line protection, 
which are responsible for tripping the line in case of instantaneous overloads.

We applied fourth order Runge-Kutta method (RK4 from Numerical Recipes)
~\cite{NumR} to solve Eq.~(\ref{11}) on various networks. Step sizes:
$\Delta = 0.1, 0.01, 0.001$ and the convergence criterion $\epsilon = 10^{-12}$ 
were used in the RK4 algorithm.
Generally the $\Delta =0.001$ precision did not improve the stability of the
solutions except at large $K$-s, while $\Delta = 0.1$ was insufficient,
so most of the results presented here are obtained using $\Delta = 0.01$.
In case of the 2D and US-HV grids we applied $\langle \omega_i^0\rangle = 0$
self-frequencies~\footnote{Due to the
Galilean invariance of Eq.~(\ref{kur2eq}) we can gauge out the mean value
in a rotating frame.}, while in case of the HU-HV the mean-values come from
the first term of right hand side of Eq.~(\ref{11}). 
For modeling uncorrelated fluctuations we added random numbers $\xi_i$ 
to the self-frequencies $\langle \omega_i^0\rangle $, following unit variance 
Gaussian distribution. To model correlated fluctuations we added $\xi_i$-s
with exponential tail distributions of the form: 
$p(\xi_i) = | \kappa \exp(-\xi_i) |$.

The initial state was fully synchronized: $\theta_i(t)=0$, $\dot{\theta_i}(t)=0$, 
but for testing the hysteresis we used uniform random distribution of phases: 
$\theta_i(t) \in (0,2\pi)$. Note, that these conditions do not correspond to
a fixed point, characterized by the sum over all flows $\sum_j F_{ij}$ 
being equal to the generated power at each node $i$. 
Thermalization was performed by running the code for $10^5$ iterations.
Following that we perturbed the system by removing a randomly selected
node in order to simulate a power failure event. After this initial node 
removal the dynamics was simulated according to Eq.~(\ref{kur2eq}) or 
Eq.~(\ref{1}) and lines are cut dynamically, according to the criterion 
(\ref{thr-cond}).
We also tried such perturbations by line cuts, but these caused too small 
cascades for making statistical analysis. 
We also tried multiple, simultaneous random node removals, which caused larger, 
but identical blackout distributions as the single node case. 
During the cascade simulations, which had the length of $t_{max}=10^4$
\footnote{Throughout the simulations we assumed dimensionless units for the
time, but in case of the HU-HV we had parameters, with real SI units, thus here
time can be interpreted with units of $s$.}
we measured the Kuramoto order parameter:
\begin{equation}\label{ordp}
z(t_k) = r(t_k) \exp{i \theta(t_k)} = 1 / N \sum_j \exp{[i \theta_j(t_k)}] \ ,
\end{equation}
by increasing the sampling time steps exponentially :
\begin{equation}
t_k = 1 + 1.08^{k} \ ,
\end{equation}
where $0 \le r(t_k) \le 1$ gauges the overall coherence and $\theta(t_k)$ is
the average phase. We solved (\ref{kur2eq}) numerically for $10^4 - 10^6$
independent initial conditions, with different $\omega_i^0$-s and
determined the sample average: $R(t_k) = \langle r(t_k)\rangle$ 
%and the variance of the frequencies $\langle \ {\mathrm var} \omega_i(t)\rangle$
. 
We also recorded the total number of line failures $N_f$ of each
sample and calculated the probability distribution $p(N_f)$ of them. 
In the steady state, which we determined by visual inspection of
the mean values, we measured the standard deviation: $\sigma_R$ of 
$R(t_k)$ in order to locate the transition point.

%%%%%%%%%%%%%%%%%%%%%%%%%%%%%%%%%%%%%%%%%%%%%%%%%%%%%%%%%%%%%%%%%%%%%%%%%
\subsection{Description and Analysis of the power-grids}
%%%%%%%%%%%%%%%%%%%%%%%%%%%%%%%%%%%%%%%%%%%%%%%%%%%%%%%%%%%%%%%%%%%%%%%%%

\begin{figure}[H]
\centering
\includegraphics[width=14cm]{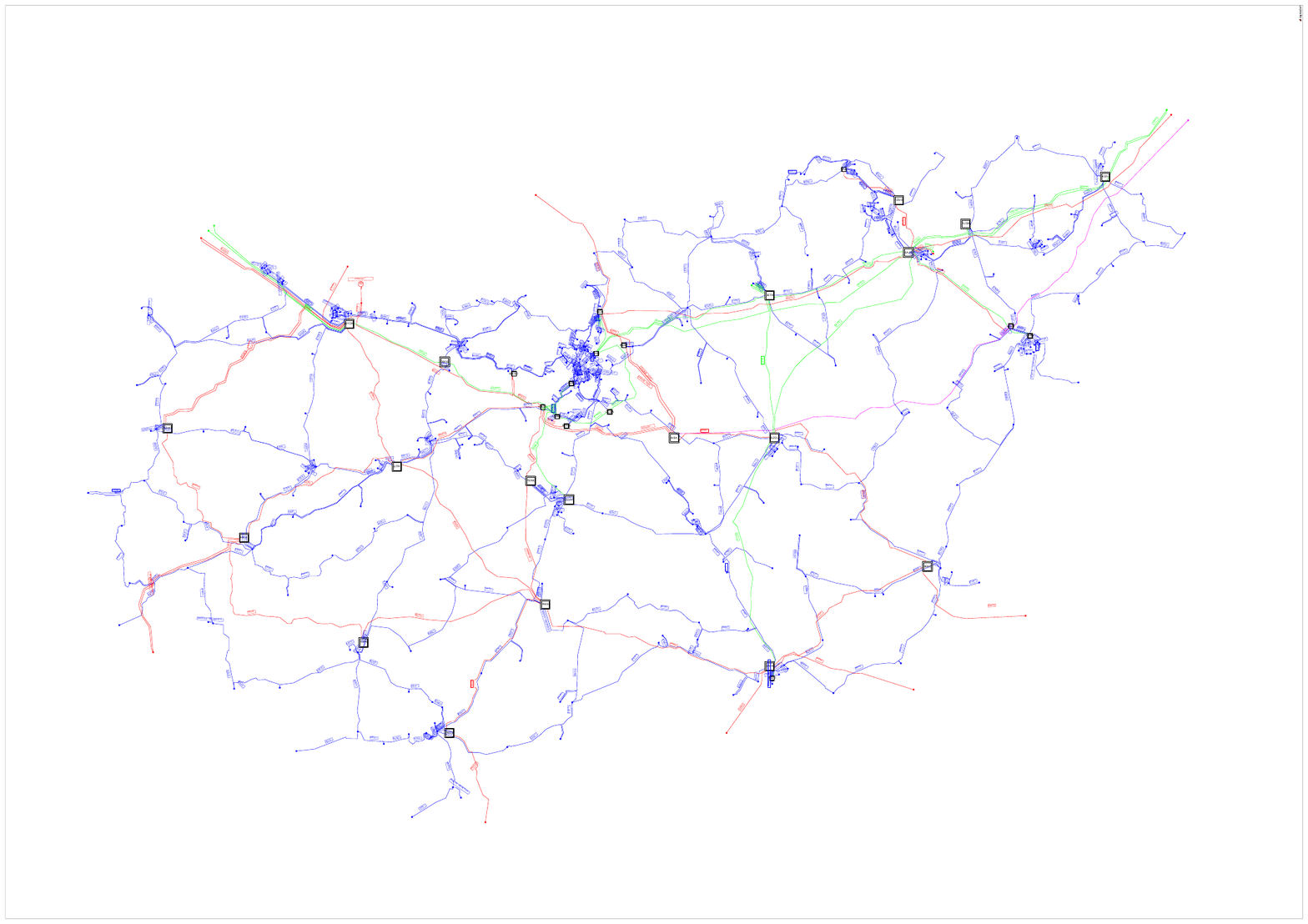}
\caption{The topography of Hungarian transmission (750 kV -- purple, 400 kV
-- red, 220 kV -- green) and sub-transmission (120 kV -- blue) networks
.\label{Fig1}}
\end{figure}

\begin{figure}[H]
\centering
\includegraphics[width=10cm]{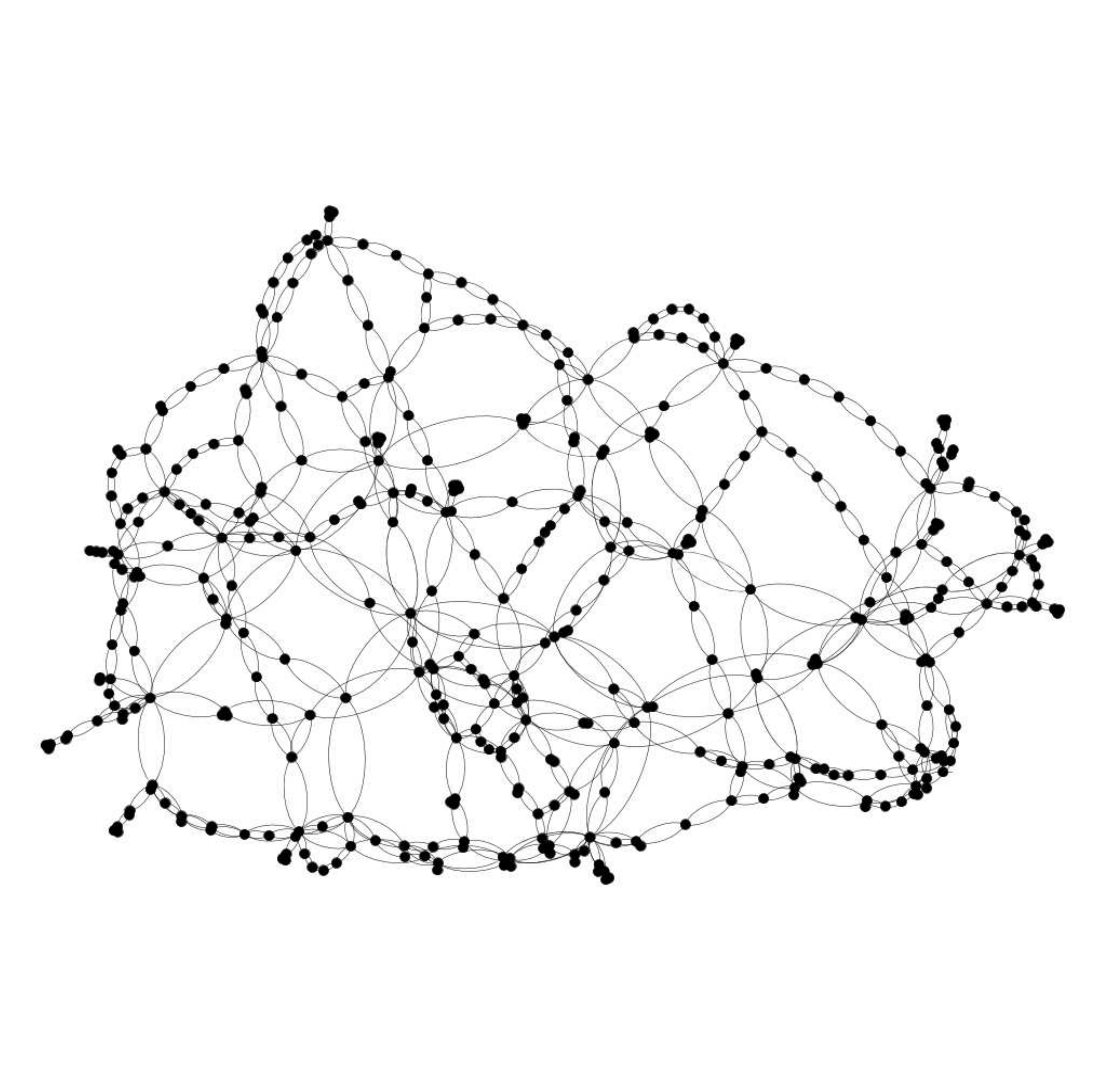}
\caption{The topology of the HU-HV grid. Note, that in transmission
networks unidirectional lines may occur, but here double lines were 
also modeled as single connections.\label{HUW1}}
\end{figure}

To create the model of the Hungarian HV power grid, the authors have
relied dominantly on the data provided by MAVIR (see Fig.~\ref{Fig1}). 
Complete topology of 750, 400, 220 transmission and
120 kV sub-transmission networks has been replicated with 418 nodes.
The topology of these systems (see Fig.~\ref{HUW1}) is mostly looped and 
meshed, with only a number of direct lines. 
The model includes approx. 50 larger power plants, 200 composite distributed 
generators, which represent units of mixed fuel (gas engines, solar 
photovoltaics, wind turbines) and 200 loads. 
The generation mix, the share of converter-based generation
units in the portfolio and the value of  ${K}_{i}$ couplings were
determined using statistics of the Hungarian Energy and Public Utility
Regulatory Authority and MAVIR, while $P_i$, and $I_i$ values were set
according to empirical distributions created from historical data.
${H}_{\mathit{Gi}}$ and  ${H}_{\mathit{Li}}$ were $5.5[s]$ and $0.36[s]$,
respectively.
%%%%%%%%%%%%%%%%%%%%%%%%%%%%%%%%%%%%%%%%%%%%%%%%%%%%%%%%%%%%%%%%%%%%%%%%%
%%%%%%%%%%%%%%%%%%%%%%%%%%%%%%%%%%%%%%%%%%%%%%%%%%%%%%%%%%%%%%%%%%%%%%%%

%%%%%%%%%%%%%%%%%%%%%%%%%%%%%%%%%%%%%%%%%%%%%%%%%%%%%%%%%%%%%%%%%%%%%%%%%
%%%%%%%%%%%%%%%%%%%%%%%%%%%%%%%%%%%%%%%%%%%%%%%%%%%%%%%%%%%%%%%%%%%%%%%%

We determined some basic topology characteristics ~\cite{Newmanbook} of 
this graph using the Gephi tool~\cite{gephi}.
The $N=418$ nodes of the network are interconnected via $E=1077$ undirected
links. The average degree is: $\langle k\rangle = 2.595$ and the
exponent of the cumulative degree distribution is: $\gamma=1.51(4)$,
which renders this network just at the threshold of robust/fragile:
$\gamma=3/2$, according to the definition by~\cite{Pgridtop}.
Note, that in the publication~\cite{CasTh} only the $220$ and $400$ kV
infrastructure of the Hungarian HV network was considered, which is
a smaller sub-network with $N=40$ , possessing more fragile geometry 
than the model used for present paper. 

The HU-HV is a highly modular network with modularity quotient
$Q=0.8$, defined by
\begin{equation}
Q=\frac{1}{N\av{k}}\sum\limits_{ij}\left(A_{ij}-
\frac{k_ik_j}{N\av{k}}\right)\delta(g_i,g_j),
\end{equation}
where $A_{ij}$ is the adjacency matrix and $\delta(i,j)$ is the Kronecker
delta function.  
The Watts-Strogatz clustering coefficient \cite{WS98} of the
network of $N$ nodes is
\begin{equation}\label{Cws}
C = \frac1N \sum_i 2n_i / k_i(k_i-1) \ ,
\end{equation}
where $n_i$ denotes the number of direct edges interconnecting the
$k_i$ nearest neighbors of node $i$, $C=0.076$ is about $10$ times 
higher, than that of a random network of same size $C_r=0.0062$,
defined by $C_r = \langle k\rangle / N$.
The average shortest path length is
\begin{equation}
L = \frac{1}{N (N-1)} \sum_{j\ne i} d(i,j) \ ,
\end{equation}
where $d(i,j)$ is the graph distance between vertices $i$ and $j$.
In case of HU-HV this is $L = 8.163$, somewhat larger than that of the
random network of same size: $L_r = 6.2244$ obtained by the formula~\cite{Fron}
\begin{equation}
L_r = \frac{\ln(N) - 0.5772}{\ln\langle k\rangle} + 1/2 \ .
\end{equation}
So, this is a small-world network, according to the definition of
the coefficient~\cite{HumphriesGurney08}:
\begin{equation}
\sigma = \frac{C/C_r}{L/L_r} \ ,
\label{swcoef}
\end{equation}
because $\sigma=9.334$ is much larger than unity.

We have also studied the dynamical behavior on the western states 
power-grid of US-HV that we downloaded from~\cite{USpg}.
This is a standard modular network, in which all transmission 
lines are bidirectional and identical, but other 
(distribution...etc) lines are omitted.
Nodes are also identical and featureless.
The network invariants are summarized in the Table~\ref{UStab}.
\begin{table}[ht]
 \caption{\label{UStab} Network invariants of the US-HV grid}
 \centering
 \begin{tabular}{|c|c|c|c|c|c|c|}
  \hline
   $N$  & $E$  & $L$  & $\langle k\rangle$ & $L_r$ & $C$ & $C_r$ \\
   \hline
   4194 & 6594 & 2.67 &         18.7       &  3.15 & 0.08 & 0.005 \\
 \hline
 \end{tabular}
\end{table}
As we can see this network is about $10$ times larger than the 
HU-HV, but exhibits similar network invariant values. 
The small world coefficient is large again: $\sigma=18.88$.
The the cumulative degree distribution is: $\gamma=1.246$, categorizing
it a robust network, by static topological sense. Later we shall
investigate if this holds in the dynamical sense, in the presence
of fluctuating energy resources.
%%%%%%%%%%%%%%%%%%%%%%%%%%%%%%%%%%%%%%%%%%%%%%%%%%%%%%%%%%%%%%%%%%%%%%%%%
\begin{figure}[H]
\centering
\includegraphics[width=10cm]{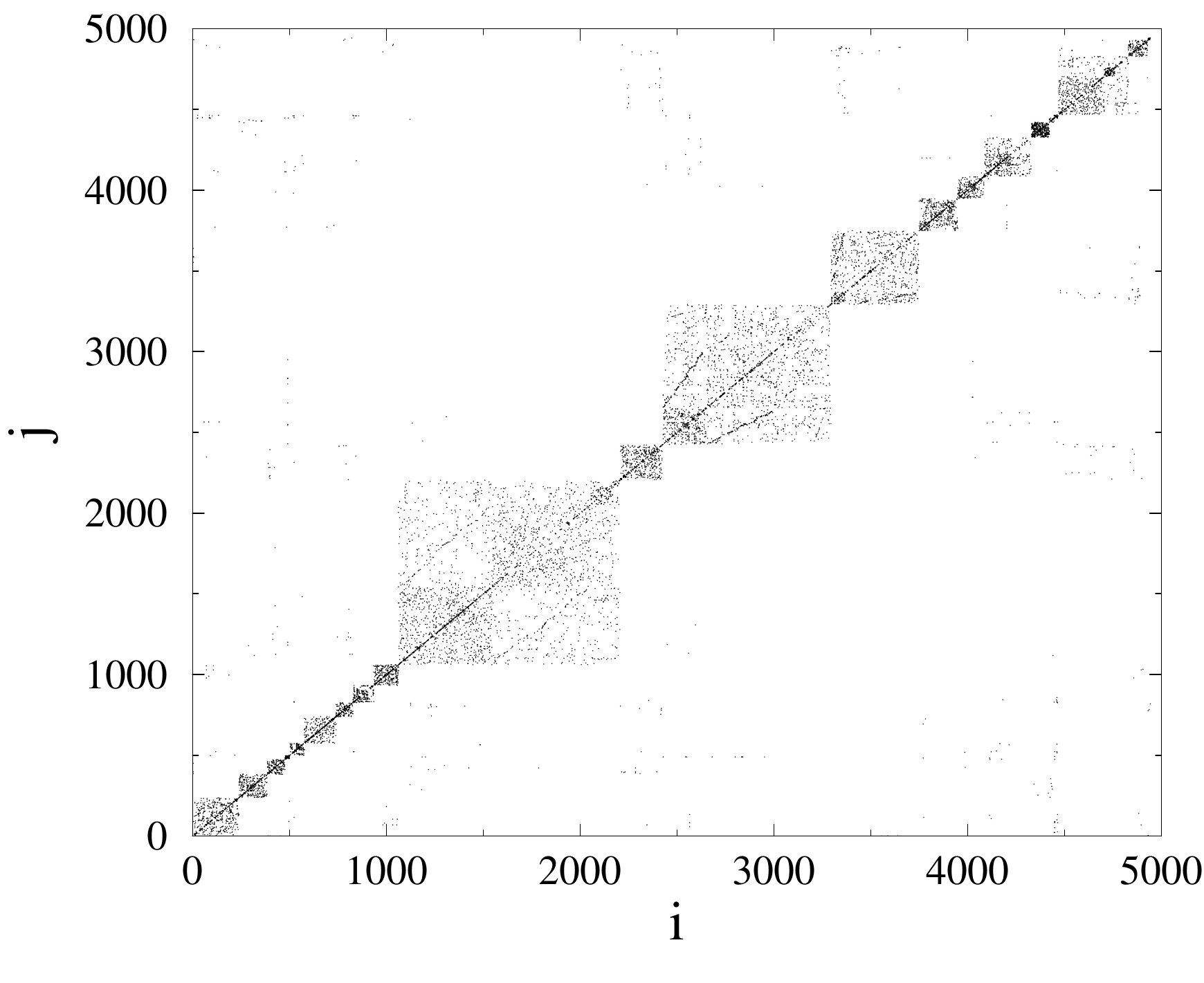}
\caption{Adjacency matrix of of the US-HV grid. Dots mark nodes $i$ and
$j$ connected.\label{USA}}
\end{figure}
%%%%%%%%%%%%%%%%%%%%%%%%%%%%%%%%%%%%%%%%%%%%%%%%%%%%%%%%%%%%%%%%%%%%%%%%

\begin{figure}[H]
\centering
\includegraphics[width=10cm]{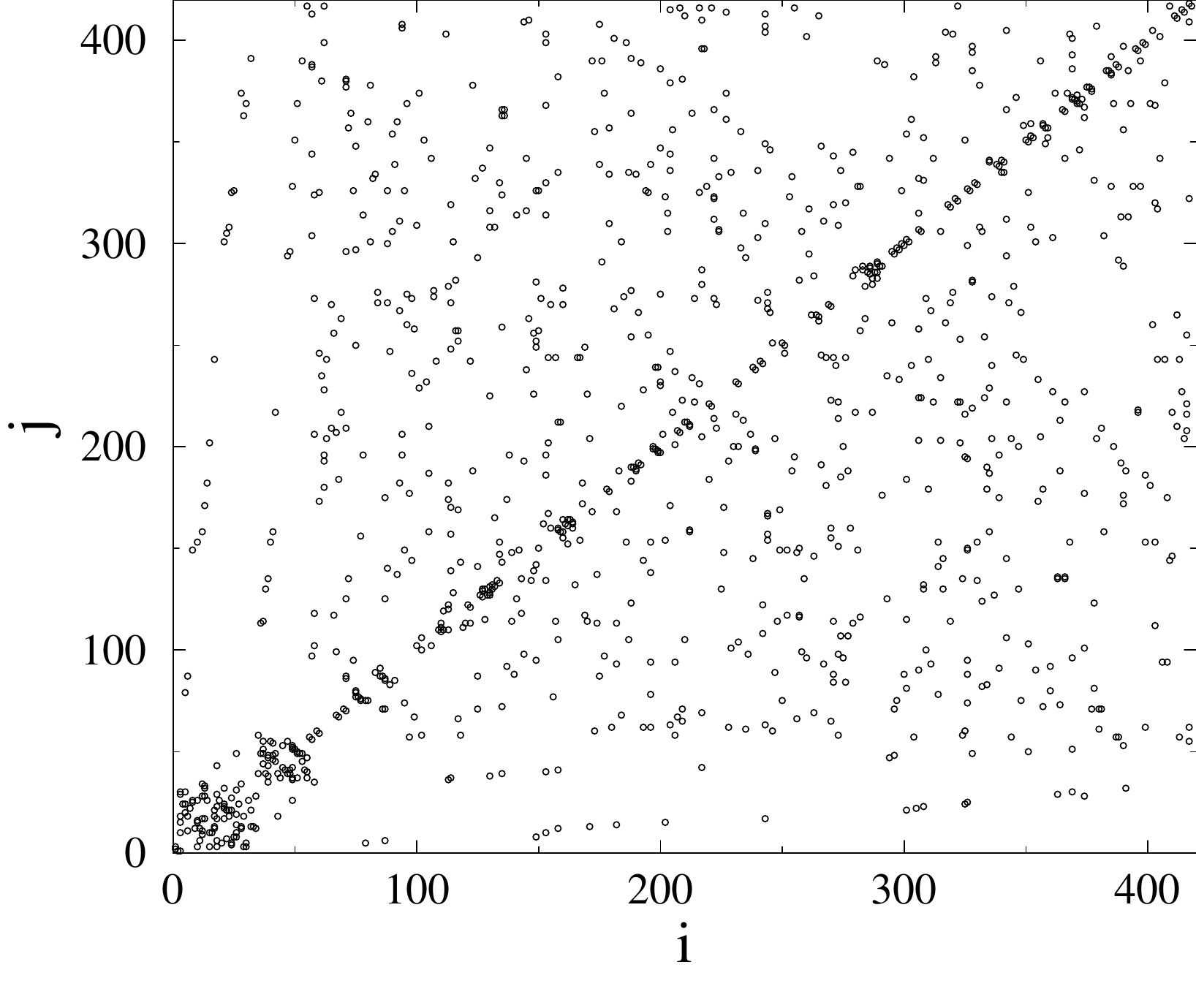}
\caption{Adjacency matrix of of the HU-HV grid. Circles mark nodes $i$ and
$j$ connected.\label{HUW1A}}
\end{figure}

By looking at the adjacency matrix of the $N=418$ node HU-HV grid 
(Fig.~\ref{HUW1A}) we can see some blocks, especially for node numbers 
$i \le 40$, corresponding to the sub-network, considered in
\cite{CasTh}, but many other connections, resembling like a random 
structure are also present. This is in contrast with the US-HV grid
(Fig.~\ref{USA}), where a more regular, HMN structure is visible.
This does not mean the lack of HMN structure of the Hungarian system
had we considered lower levels~\cite{POWcikk}, but suggests a more
random-like structure. Note, that in ref.~\cite{CasTh} more 
random-like structures were found to be more robust.
%%%%%%%%%%%%%%%%%%%%%%%%%%%%%%%%%%%%%%%%%%%%%%%%%%%%%%%%%%%%%%%%%%%%%%%%%
%%%%%%%%%%%%%%%%%%%%%%%%%%%%%%%%%%%%%%%%%%%%%%%%%%%%%%%%%%%%%%%%%%%%%%%%

%%%%%%%%%%%%%%%%%%%%%%%%%%%%%%%%%%%%%%%%%%%%%%%%%%%%%%%%%%%%%%%%%%%%%%%%%
\section{Simulation results}
%%%%%%%%%%%%%%%%%%%%%%%%%%%%%%%%%%%%%%%%%%%%%%%%%%%%%%%%%%%%%%%%%%%%%%%%%

In this section we will compare the results of the threshold Kuramoto
simulations on different networks, by gradually increasing the spatial
heterogeneity. We start from the homogeneous two-dimensional square lattice,
in which only the self-frequencies at different nodes vary randomly. 
Then we move on to the standard US-HV power grid, possessing topological 
heterogeneity too.
Finally, we consider the most realistic HU-HV, in which even the edges and
node parameters change. We supplement the HU-HV case with a feedback
control study as well.  

%%%%%%%%%%%%%%%%%%%%%%%%%%%%%%%%%%%%%%%%%%%%%%%%%%%%%%%%%%%%%%%%%%%%%%%%%
\subsection{The two-dimensional square lattice}
%%%%%%%%%%%%%%%%%%%%%%%%%%%%%%%%%%%%%%%%%%%%%%%%%%%%%%%%%%%%%%%%%%%%%%%%

\begin{figure}[H]
\centering
\includegraphics[width=10cm]{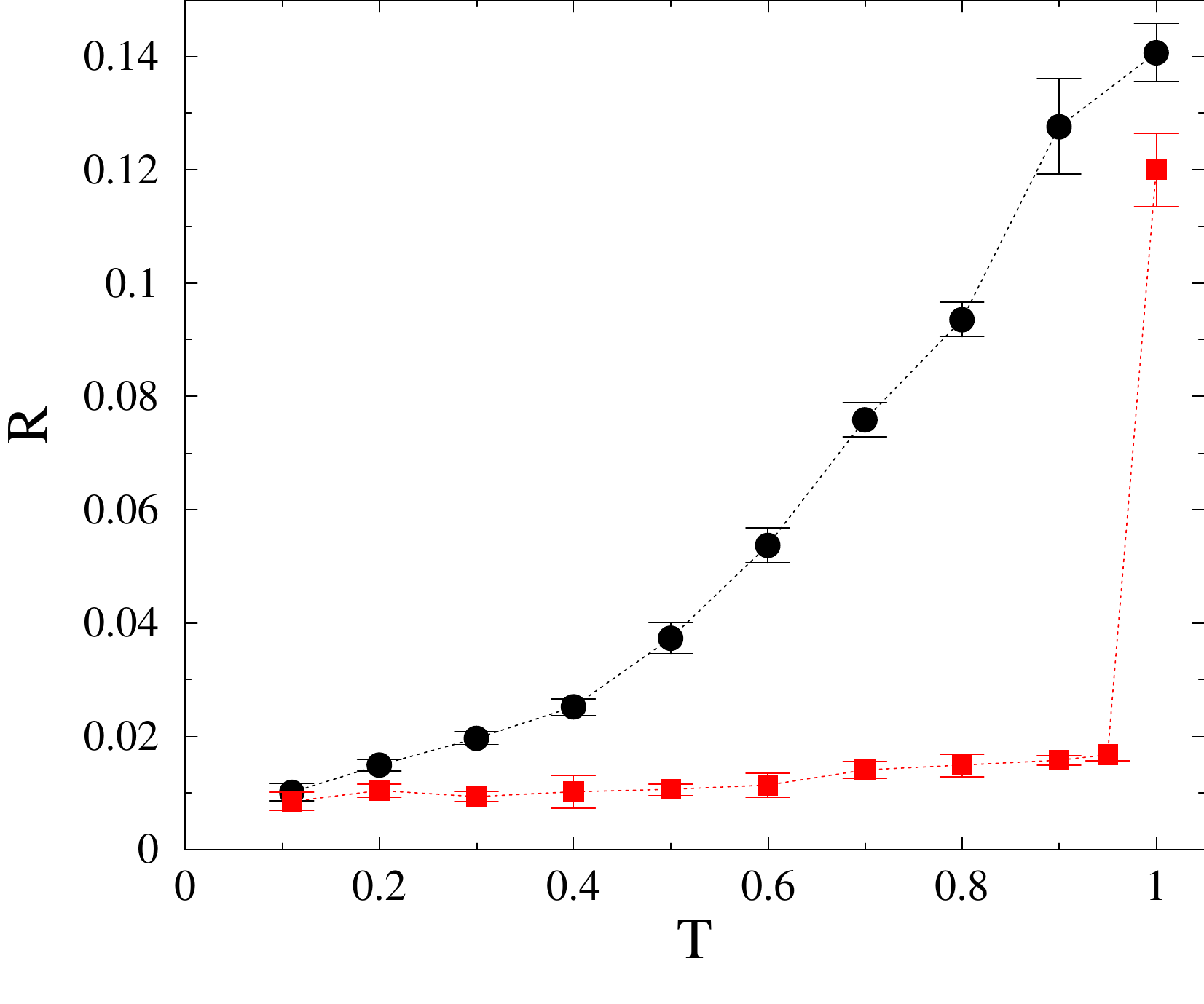}
\caption{\label{2Dss} Hysteresis of the Kuramoto order parameter on 
the 2D square lattice as the function of threshold at $K=10$.
Upper branch bullets corresponds to synchronized initial state, while 
lower branch boxes to random initialization of $\theta_i$.}
\end{figure}

\begin{figure}[H]
\centering
\includegraphics[width=10cm]{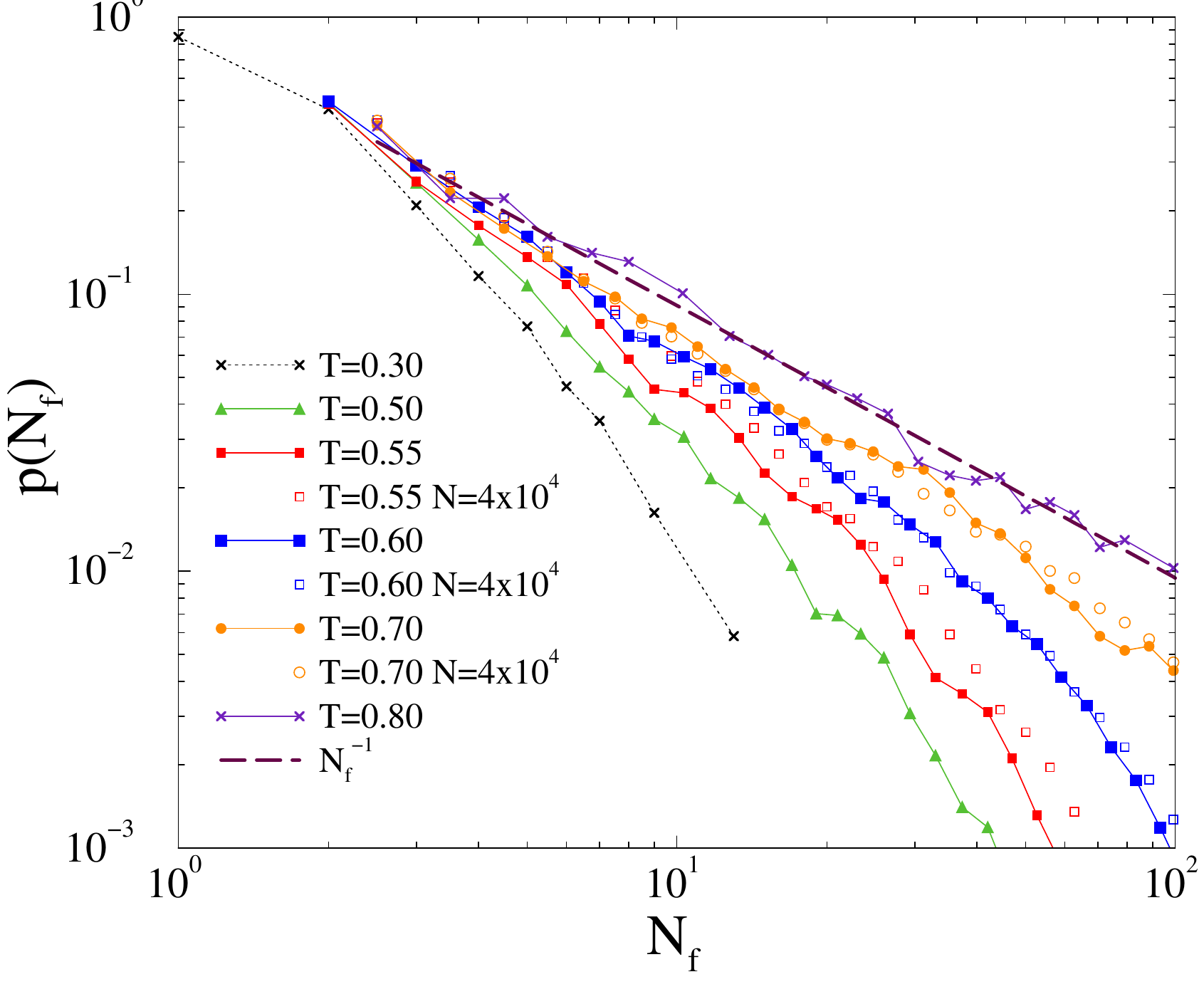}
\caption{Probability distributions of line failures for different failure 
thresholds ($T$), shown by the legends, at $K=10$. 
Closed symbols: $N=10^4$, open symbols: $N=4\times 10^4$. 
The dashed line shows a PL fit for the $T=0.8$ data, 
corresponding to the singular $p(N_f) \simeq 1/N_f$ case, lying 
in the disordered phase.
\label{2D-tail}}
\end{figure}

To determine the consequences of topological heterogeneity we have run the 
analysis using Eq.~(\ref{kur2eq}) on $N=10^4$ and  $N=4\times 10^4$ sized 
lattices, with periodic boundary conditions. 
Here we found signatures of first order synchronization transitions 
with wide hysteresis loops (see Fig.\ref{2Dss}). This is similar to
the results we obtained for the 2D second order Kuramoto in 
Ref.~\cite{POWcikk} without allowing line failures. The hysteresis means
a difficulty of the restoration of the synchronous state following a
blackout collapse.
%%%%%%%%%%%%%%%%%%%%%%%%%%%%%%%%%%%%%%%%%%%%%%%%%%%%%%%%%%%%%%%%%%%%%%%%%
%%%%%%%%%%%%%%%%%%%%%%%%%%%%%%%%%%%%%%%%%%%%%%%%%%%%%%%%%%%%%%%%%%%%%%%%
Synchronization transition is visible clearly at lower global coupling 
values only. 
For $K > 10$ the transition becomes smooth, the system remains mostly 
in the partially synchronized state.
There are no signatures of PL-s in the Kuramoto order parameter $R(t)$ curves, 
they converge quickly to their steady state values for all $K$ values.
The distribution of the total number of line failures also do not exhibit PL-s, 
but break down exponentially, or follow the singular $p(N_f) \simeq 1/N_f$ 
behavior, corresponding to the synchronization state for $K > K_c \simeq 0.7$ 
before the finite size cutoff (see Fig.~\ref{2D-tail}).
%%%%%%%%%%%%%%%%%%%%%%%%%%%%%%%%%%%%%%%%%%%%%%%%%%%%%%%%%%%%%%%%%%%%%%%%%
%%%%%%%%%%%%%%%%%%%%%%%%%%%%%%%%%%%%%%%%%%%%%%%%%%%%%%%%%%%%%%%%%%%%%%%%
By increasing the system size from $N=10^4$ to $N=4\times 10^4$ the 
results did not change, as shown in the figure for the $K=0.55,0.6,0.7$ 
coupling cases.
Note, that the average size of the blackouts decrease with $T$, because
several links are already removed during the thermalization process before
the actual cascade simulations started. 

%%%%%%%%%%%%%%%%%%%%%%%%%%%%%%%%%%%%%%%%%%%%%%%%%%%%%%%%%%%%%%%%%%%%%%%%%
\subsection{The US-HV Power-grid}
%%%%%%%%%%%%%%%%%%%%%%%%%%%%%%%%%%%%%%%%%%%%%%%%%%%%%%%%%%%%%%%%%%%%%%%%

\begin{figure}[H]
\centering
\includegraphics[width=10cm]{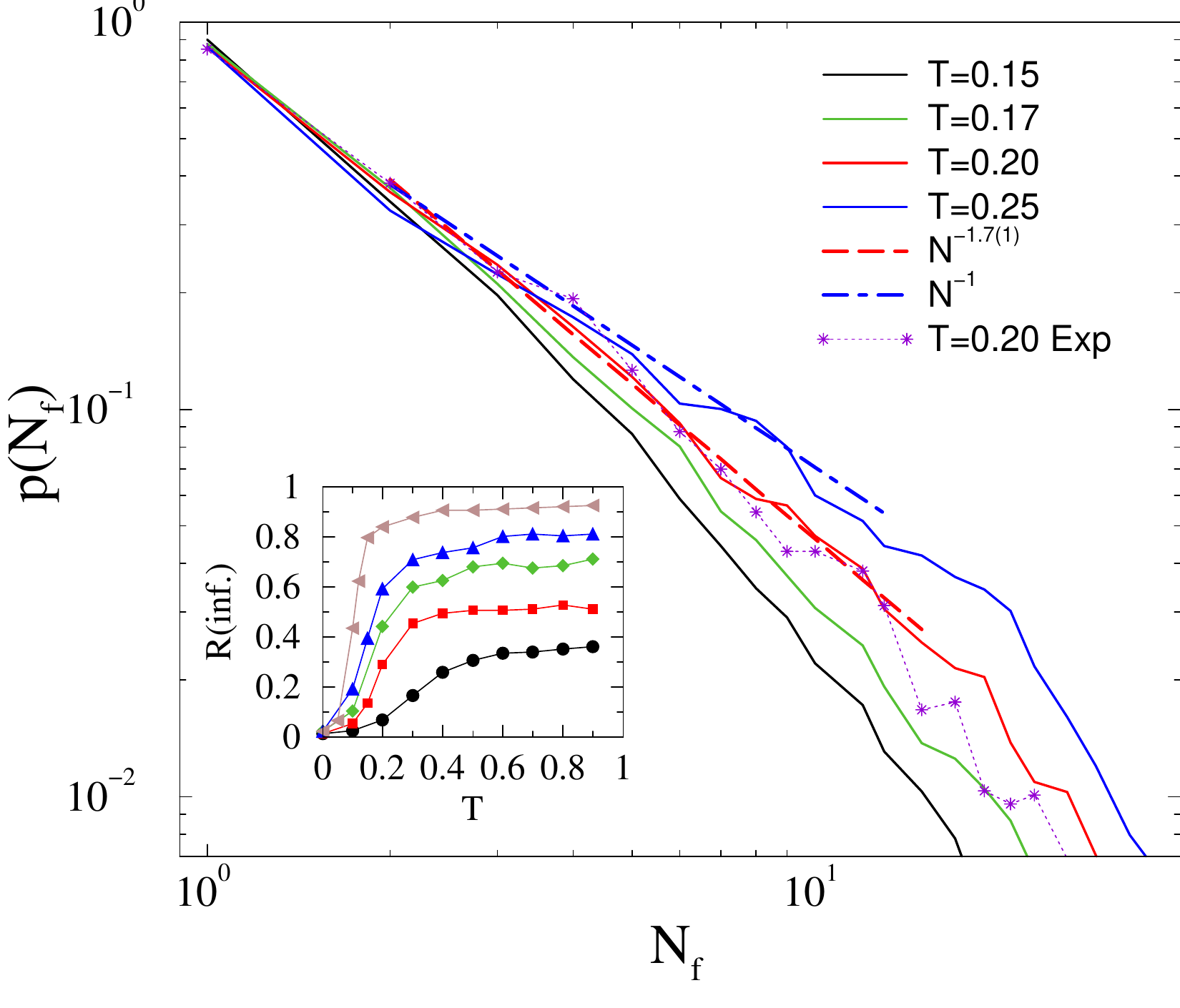}
\caption{Probability distribution of line failures for different thresholds
at $K=30$ shown in the legends in case of the US power-grid.
Lines corresponds to Gaussian distributed $\omega_i^0$-s,
while star symbols to the exponentially distributed self-frequencies 
in case of $T=0.2$. Dashed lines show power-law fits for the scaling region, 
determined by visual inspection.
The inset shows $R(t\to\infty)$ as the function of time, for 
$K=10,20,30,40,70$ (bottom to top curves).
\label{US-tail}}
\end{figure}

\begin{figure}[H]
\centering
\includegraphics[width=10cm]{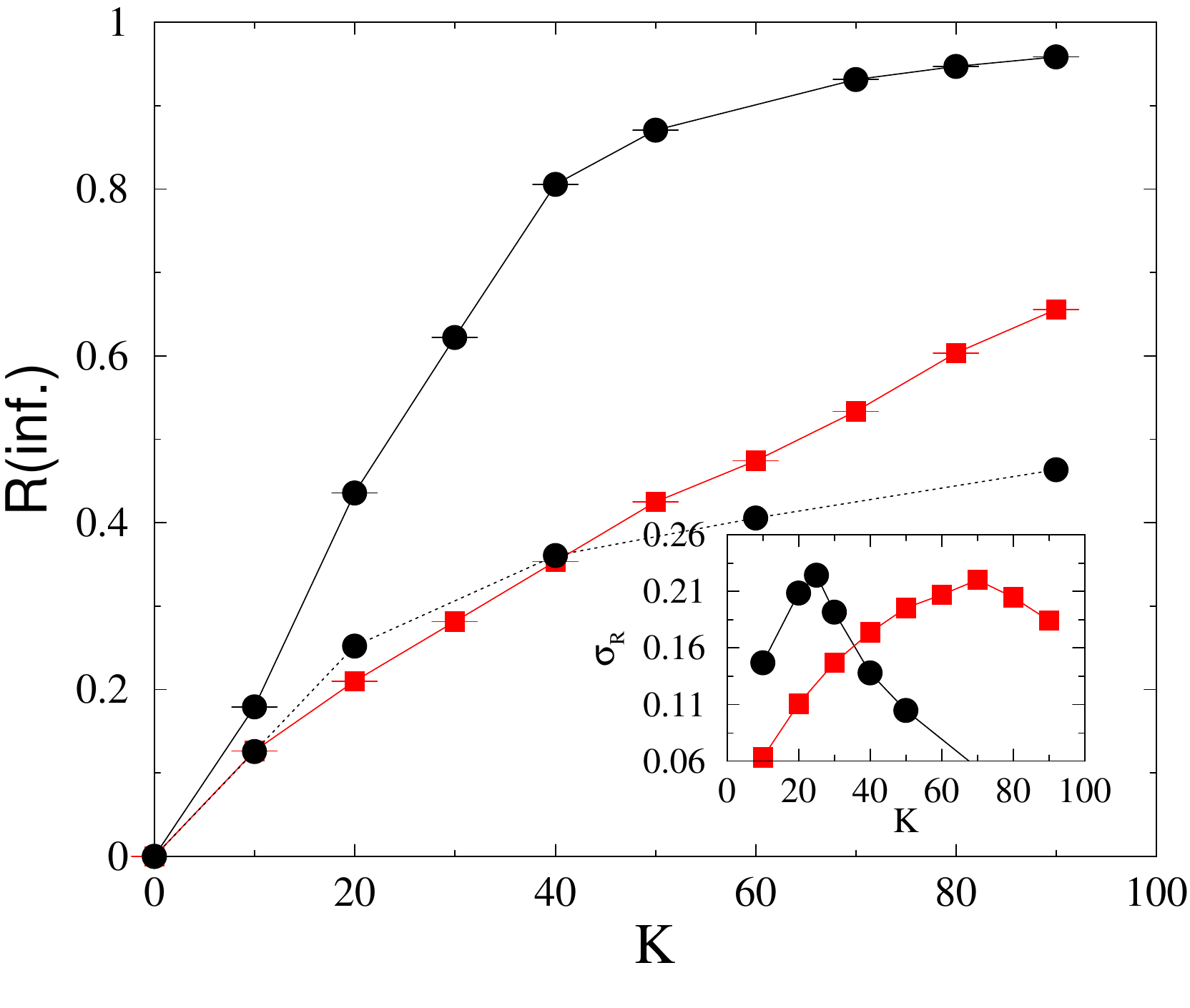}
\caption{Steady state order parameter as the function of $K$ for $T=0.3$ 
in case of the US-HV. Black bullets are for Gaussian, while red boxes are for
exponential tailed $g(\omega_i^0)$ self-frequency distributions. The two
branches of Gaussian correspond to ordered and disordered initial states
representing a hysteresis loop, closing at $K>400$.
The inset shows the fluctuations, $\sigma_R$ of the same.
\label{betaT03}}
\end{figure}

Next we performed dynamical simulations using Eq.(~\ref{kur2eq})
on the US-HV power grid, which has topological heterogeneity as well, 
but the lines and nodes are identical.
As in case of 2D and the US-HV without line failures~\cite{POWcikk} we found
smooth crossover from desynchronization to partial synchronization by
increasing the global coupling $K$. On the other hand, there is a sudden 
jump by increasing the threshold from $T=0$ to small values. 
The inset of Fig.~\ref{US-tail} summarizes the steady state values for various
$K$-s as the function of threshold $T$. We can find a transition region 
for $T < 0.5$, that we shall investigate in more detail. 
On Fig.~\ref{betaT03} we show the steady state behavior at fixed $T=0.3$ 
as the function of $K$. 
At this threshold the fluctuation peak $\sigma_R$ marks a transition 
point at $K\simeq 25$.
One can also see the lower part of a hysteresis loop, closing at $K>400$, 
corresponding to synchronous and asynchronous initial conditions.
In case of exponential tailed $g(\omega_i^0)$ the Kuramoto order parameter
decreases and the transition point shifts to larger coupling $K\simeq 70$.

We have also investigated the dynamical behavior at $K=30$, near the transition point. 
As Fig.~\ref{US-tail} shows for Gaussian $p(\omega_i^0)$-s we find PL tailed 
$p(N_f)$ line failure distribution at $T=0.2$, which can be fitted by 
$N_{f}^{-1.7(1)}$, in agreement with the empirical data and simulations by 
Ref.~\cite{Car,Car2}. However, this PL breaks down rather early, for
$N_f < 30$, due to the finite size of the network. Another PL: 
$N_{f}^{-1}$ can be fitted for the $T=0.25$ curve, but this corresponds 
to a singular distribution, corresponding to the disordered phase, where any
kind of large cascade may occur, restricted by the finite grid size. 
%%%%%%%%%%%%%%%%%%%%%%%%%%%%%%%%%%%%%%%%%%%%%%%%%%%%%%%%%%%%%%%%%%%%%%%%%
%%%%%%%%%%%%%%%%%%%%%%%%%%%%%%%%%%%%%%%%%%%%%%%%%%%%%%%%%%%%%%%%%%%%%%%%

%%%%%%%%%%%%%%%%%%%%%%%%%%%%%%%%%%%%%%%%%%%%%%%%%%%%%%%%%%%%%%%%%%%%%%%%%
%%%%%%%%%%%%%%%%%%%%%%%%%%%%%%%%%%%%%%%%%%%%%%%%%%%%%%%%%%%%%%%%%%%%%%%%

By changing the Gaussian $p(\omega_i^0)$-s to an exponential tailed one 
we cannot see difference in the line failure distribution at $T=0.2$ 
as shown on Fig.~\ref{US-tail}. 
Of course in the steady state the synchronization
drops substantially as demonstrated on Fig.~\ref{betaT03}.
Note, that a more realistic US-HV power-grid, containing node and 
line heterogeneity data would be needed to make a comparison with
real life. In the lack of this we now turn towards the Hungarian HV 
power-grid, for which we could access these data, although for a 
smaller network now. Still a comparison, in which we gradually increase 
the heterogeneity from 2D across US to HU power grids provides a useful
insight into the effects of heterogeneity on the synchronization behavior
of these models. 

%%%%%%%%%%%%%%%%%%%%%%%%%%%%%%%%%%%%%%%%%%%%%%%%%%%%%%%%%%%%%%%%%%%%%%%%%
\subsection{The Hungarian HV Power-grid}
%%%%%%%%%%%%%%%%%%%%%%%%%%%%%%%%%%%%%%%%%%%%%%%%%%%%%%%%%%%%%%%%%%%%%%%%

Next, we studied Eq.~(\ref{11}) on the empirical HU-HV power-grid, 
deduced from the Hungarian database of MAVIR. 
At first, inertia constants of nodes with purely load connections 
were set to  ${H}_{i}=0.36 s$, but very low-level synchronization was 
obtained even for $T = 1$. This is the consequence of high-level 
heterogeneity destroying the synchronization.
Thus, we modified the model by equalizing inertia constants as
${H}_{i}=5.5 s$ for most of the nodes. Exceptions were nodes with purely
generation connections, where inertia constants were selected based on
Table~\ref{tab:InertiaG} and cross-border connections, where inertia 
constants reflect different composition of generation portfolio in 
neighboring countries (ranging from $2.25 s$  to $4.5 s$).

\begin{figure}[H]
\centering
\includegraphics[width=10cm]{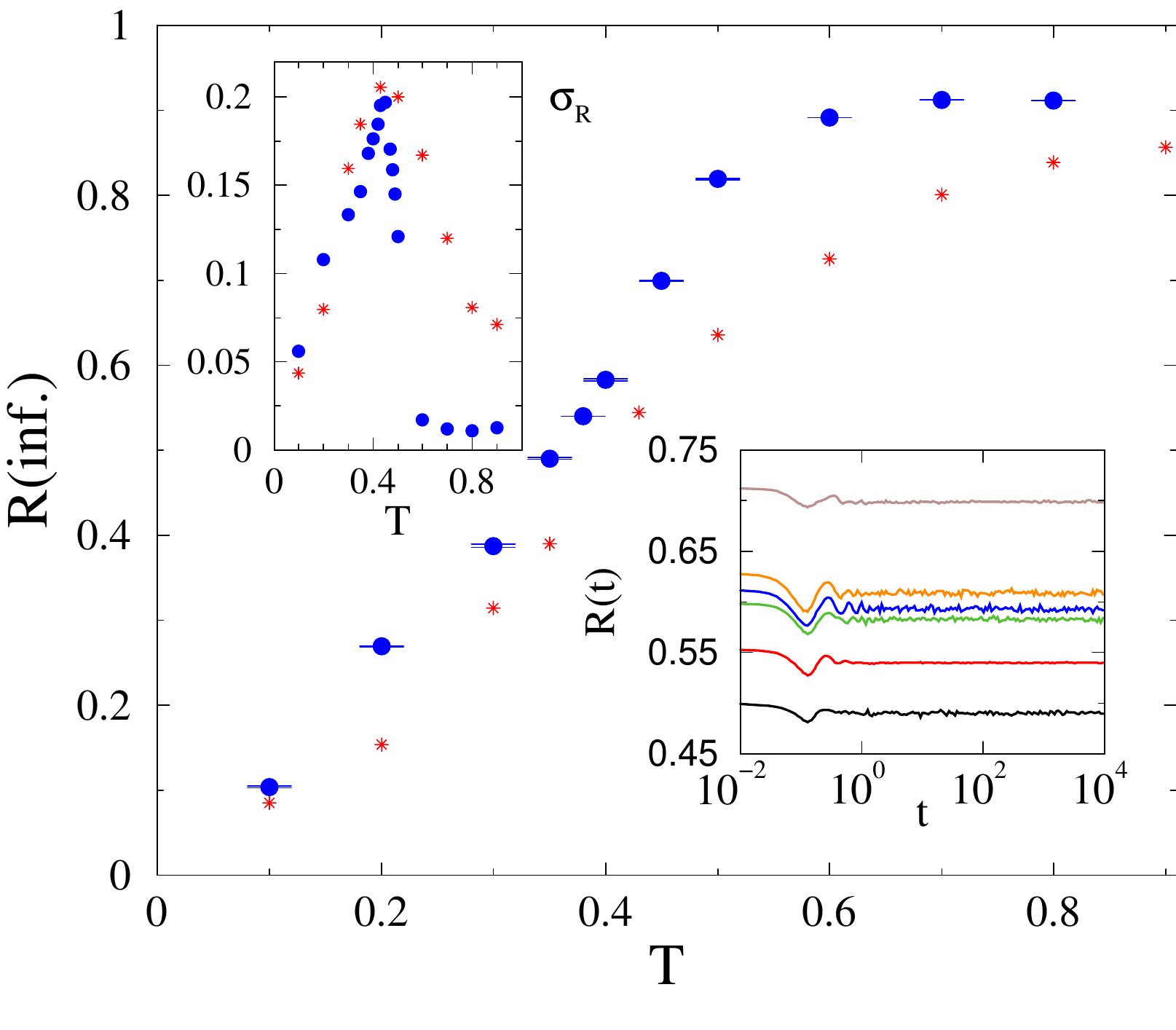}
\caption{Kuramoto order parameter in the HU-HV power-grid as
the function of threshold, bullets: Gaussian $g(\omega_i^0)$,
stars: exponential tailed fluctuations.
The upper inset shows $\sigma_R$ of the same.
The lower inset shows the time dependence in case of
$g(\omega_i^0)$ at $T = 0.35, 0.38, 0.40, 0.42, 0.43, 0.45$
(bottom to top curves).
\label{HU3-R}}
\end{figure}

Now we could find reasonable average order parameters and a synchronization 
transition as shown in Fig.~\ref{HU3-R}. The peak of the standard deviations 
of the order parameter marks a transition point at $T_c=0.44(1)$.
If we replace $g(\omega_i^0)$ from Gaussian to exponential tailed 
self-frequencies the order parameter decreases and $\sigma_R$ increases, 
but the peak does not move a lot.
%%%%%%%%%%%%%%%%%%%%%%%%%%%%%%%%%%%%%%%%%%%%%%%%%%%%%%%%%%%%%%%%%%%%%%%%%
%%%%%%%%%%%%%%%%%%%%%%%%%%%%%%%%%%%%%%%%%%%%%%%%%%%%%%%%%%%%%%%%%%%%%%%%

\begin{figure}[H]
\centering
\includegraphics[width=10cm]{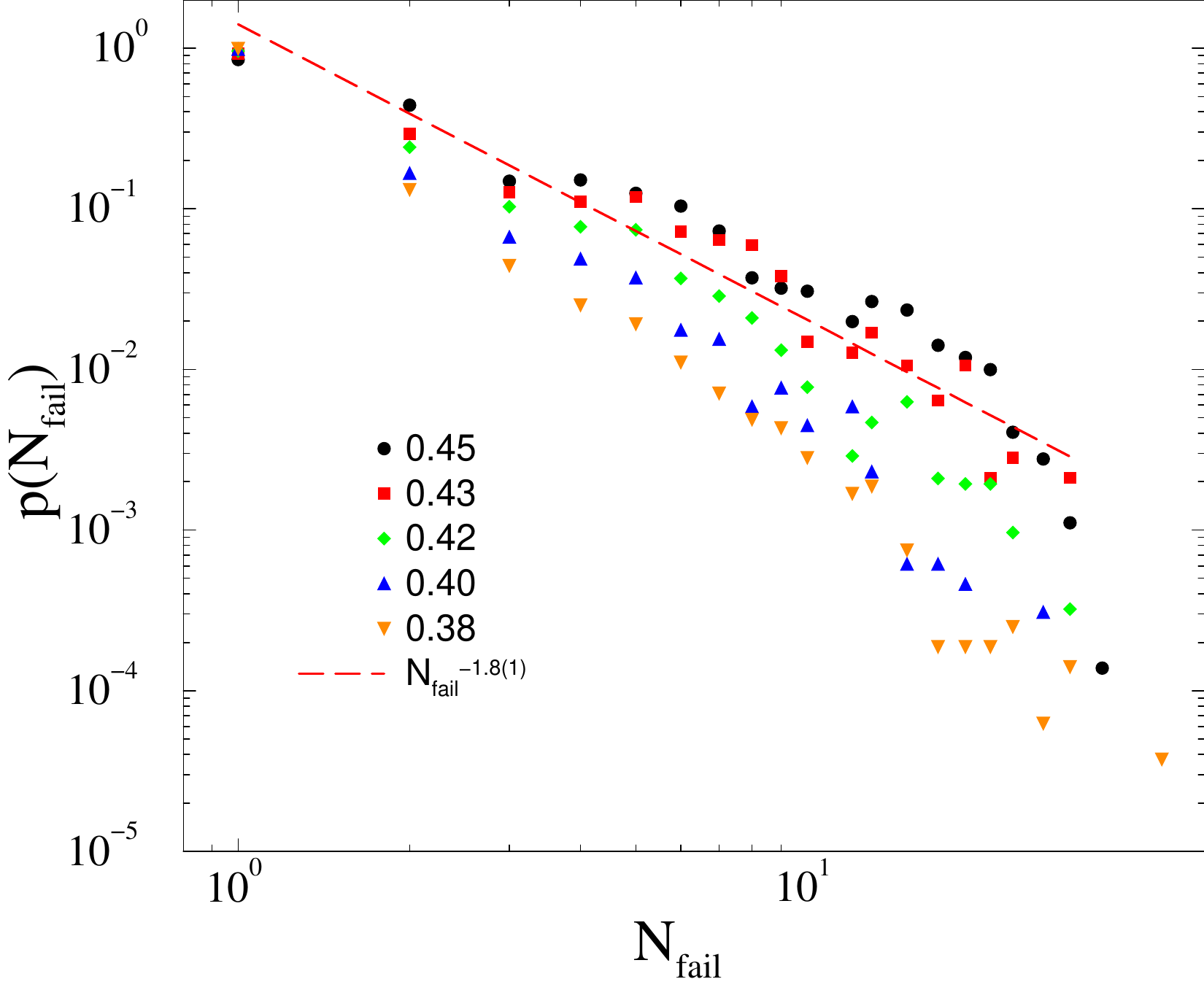}
\caption{Probability distribution of line failures for different thresholds
as shown in the legends in case of the HU-HV power-grid. The dashed line shows
a power-law fit for scaling region of the $T=0.43$ results.
\label{HU3-tail}}
\end{figure}

The probability distribution of line failures exhibit PL behavior tails 
at $T_c=0.43$, characterized by the exponent $\tau_N \ge 1.8(1)$, 
close to the blackout failure exponent as shown on Fig.~\ref{HU3-tail}.
Below the transition is hard to determine if other PL-s with 
cutoffs or a simple exponential decay happens given the small
system sizes. We favor the former scenario, but plan to test it
in the future, when larger power-grids and more computation 
resources will be at our disposal. Note, that the load dependent
PL exponents have been also advanced in case of DC threshold models
of power grids~\cite{Bis}.
Later we will investigate, if a feedback mechanism can stabilize the 
synchronization of the model with fully heterogeneous inertia.
Such feedback is present in real system, so it is an important
issue to investigate.

%%%%%%%%%%%%%%%%%%%%%%%%%%%%%%%%%%%%%%%%%%%%%%%%%%%%%%%%%%%%%%%%%%%%%%%%%
%%%%%%%%%%%%%%%%%%%%%%%%%%%%%%%%%%%%%%%%%%%%%%%%%%%%%%%%%%%%%%%%%%%%%%%%

\begin{figure}[H]
\centering
\includegraphics[width=10cm]{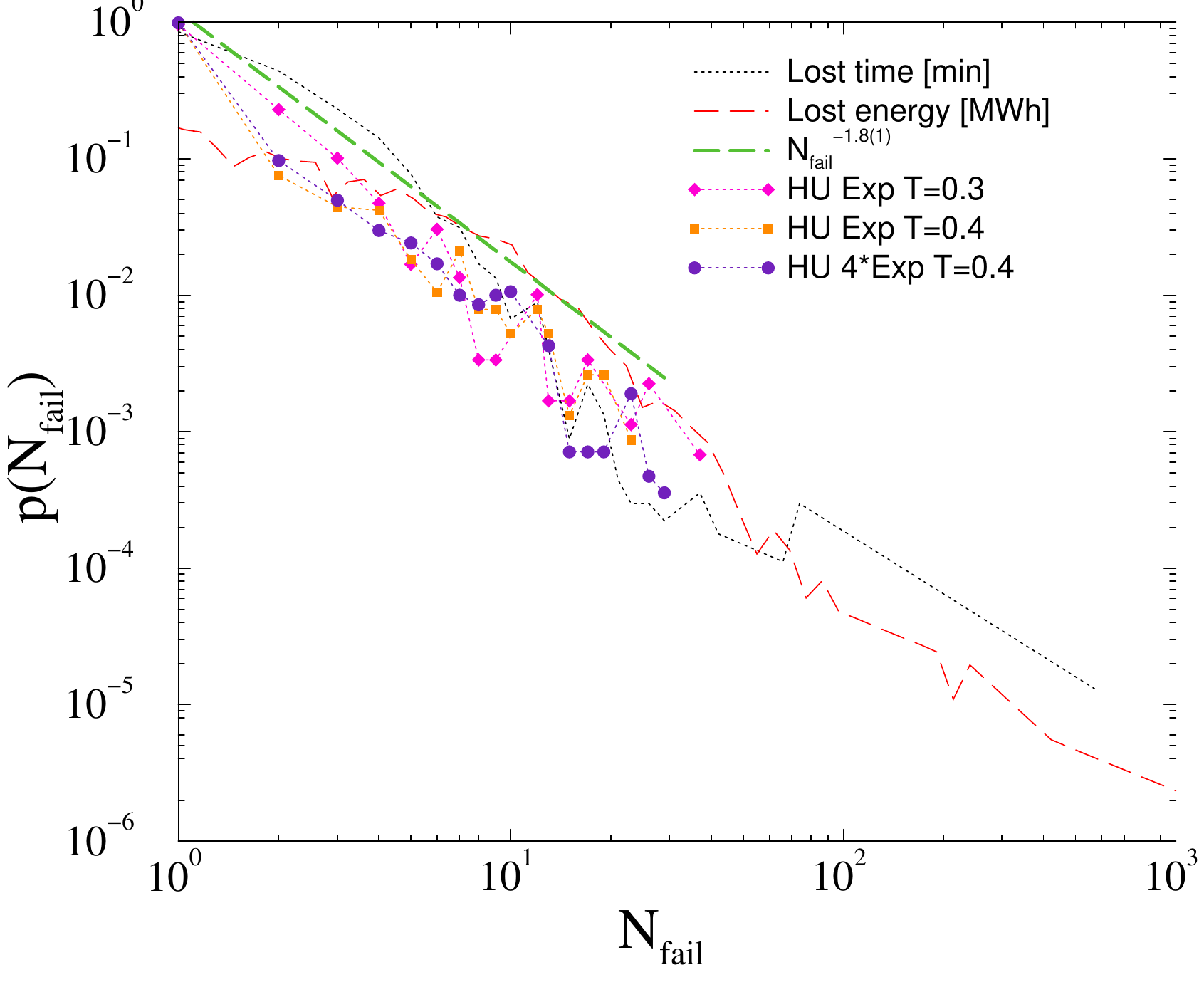}
\caption{The same as in Fig.~\ref{HU3-tail}, in case of exponential
tailed self-frequency fluctuations. The green dashed line shows a power-law 
fit for the scaling region of the $T=0.4$ threshold result shifted up for
better visibility.
For comparison we also show empirical distributions for the lost time
(black dots) and lost energy (orange dashed line)
obtained from the MAVIR database.
\label{HU3-exp}}
\end{figure}

The line failure distributions of the HU-HV power-grid seems to be quite 
insensitive for replacing the Gaussian self-frequency fluctuations to 
exponential ones. As Fig.~\ref{HU3-exp} shows the $p(N_f)$ distributions decay with 
the same PL tails as before, characterized by the exponent
$\tau_N \ge 1.8(1)$ even up to $\kappa=4$ amplitudes. Therefore,
the HU-HV power-grid model seems to be robust against large fluctuations.
%%%%%%%%%%%%%%%%%%%%%%%%%%%%%%%%%%%%%%%%%%%%%%%%%%%%%%%%%%%%%%%%%%%%%%%%
%%%%%%%%%%%%%%%%%%%%%%%%%%%%%%%%%%%%%%%%%%%%%%%%%%%%%%%%%%%%%%%%%%%%%%%%

For completeness we also show a comparison of our model calculations with
the lost time [min] and rescaled lost power [MW], obtained from planned
and unplanned outages of the Hungarian HV networks. 
The metric, described by the curve "lost energy" is also known as energy not
served (ENS), a widely accepted fundamental index of power system
reliability. ENS is defined as the expected amount of energy not being
served to consumers by the system during the period considered due to system
capacity shortages or unexpected severe power outages. Statistics of the
Hungarian transmission system were used to determine the probability
distribution of this metric. Using the same dataset, for each outage event,
we determined the amount of time that was necessary to restore operation:
this is shown by the curve "lost time".
Following appropriate rescaling we can see remarkable agreement of the
probability distributions with those obtained by our simulations.

%%%%%%%%%%%%%%%%%%%%%%%%%%%%%%%%%%%%%%%%%%%%%%%%%%%%%%%%%%%%%%%%%%%%%%%%%
\subsection{Instantaneous feedback control on the HU-HV Power-grid}
%%%%%%%%%%%%%%%%%%%%%%%%%%%%%%%%%%%%%%%%%%%%%%%%%%%%%%%%%%%%%%%%%%%%%%%%

\begin{figure}[H]
\centering
\includegraphics[width=10cm]{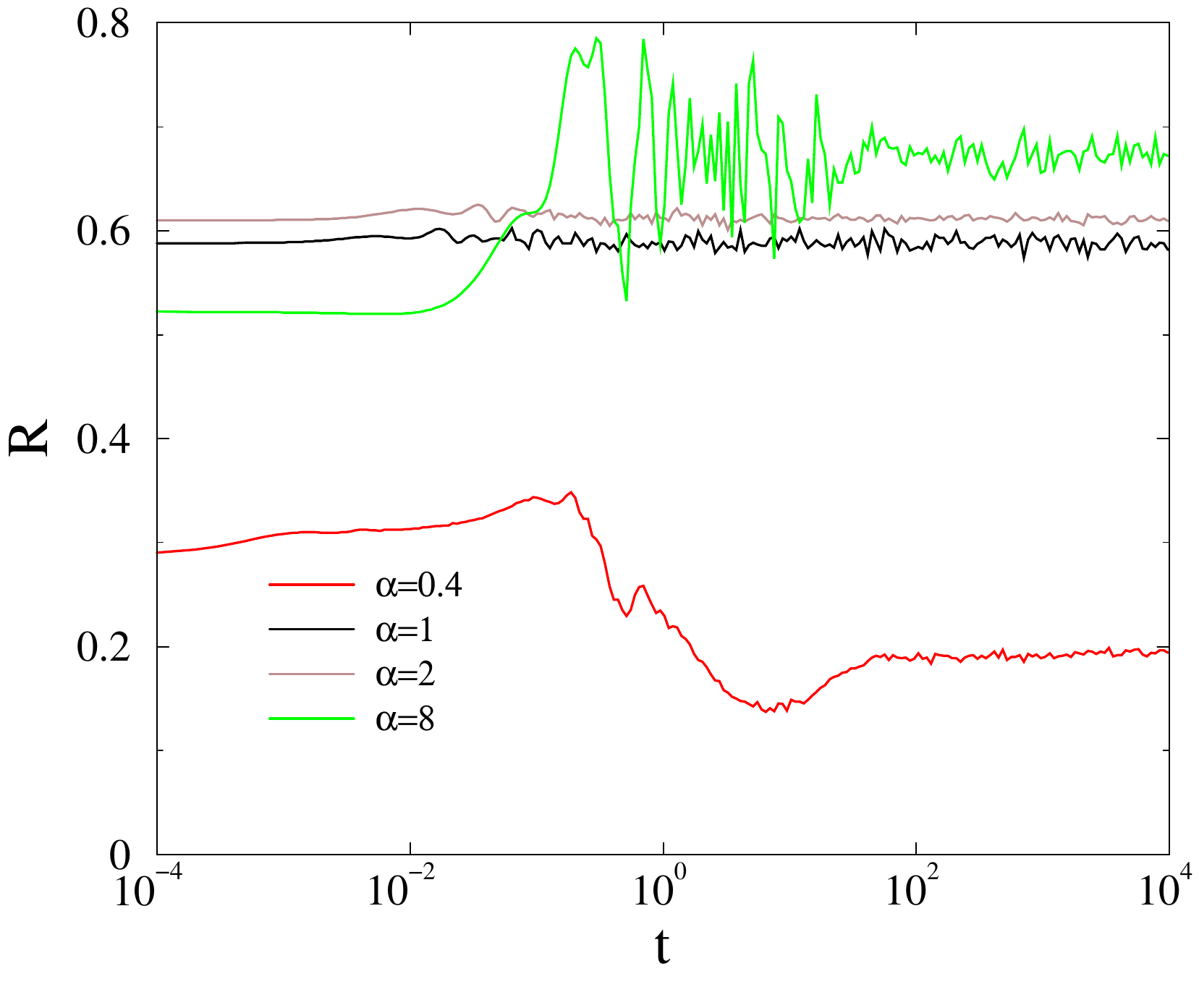}
\caption{Effect of the instantaneous feedback by increasing 
$\alpha = 0.4, 1, 2, 8$ in the HU-HV power-grid with 
heterogeneous inertia $H_i$ at $T=0.43$. 
\label{feedback}}
\end{figure}

As we mentioned, the application of Eq.~(\ref{11}) on the HU-HV
power-grid with real inertia constants shown in 
Tables~\ref{tab:InertiaG},~\ref{tab:InertiaC} leads to low
synchronization levels, the strong heterogeneity prevents to achieve
realistic synchronization values. In the recent study 
by~\cite{Olmi-Sch-19} the effects of different feedback control mechanisms
have been compared. It was shown that time delayed feedback provide
efficient ways to improve synchronization, but an instantaneous feedback
can also make the system more stable. Without going into the details
of such analysis, which is out of the scope of our present interest
we just show how an instantaneous feedback alters our results.
This can done be rather easily, since the equation of motion is
almost like the original one: Eq.~(\ref{kur2eq}):
\begin{equation}\label{kur2eqF}
\dot{\omega_i}(t) = \omega_i - \alpha \dot{\theta_i}(t) 
+ K \sum_{j=1}^{N} A_{ij} \sin[ \theta_j(t)- \theta_i(t)] 
- g \alpha \dot{\theta_i}(t) \,
\end{equation}
with the addition of a new term, describing the feedback with
gain value $g$. This can be fused with the dissipation term 
$\alpha \dot{\theta_i}(t)$, thus modeling a simple instantaneous 
feedback means enhancement of $\alpha$ in our simulations. 
Figure.~\ref{feedback} shows the time dependence of the order parameter 
by increasing $\alpha$ in case of the HU-HV power-grid model using
real, heterogeneous $H_i$ values from 
Tables~\ref{tab:InertiaG},~\ref{tab:InertiaC}.
As we can see this feedback mechanism increases $R(t)$, but the precise 
solution requires much smaller step sizes due to the high amplitudes 
of the derivatives by the integration steps. 
On Fig.~\ref{feedback} we showed $R(t)$ results using $\delta=0.0001$ 
precision, averaged over $500$ samples, because even $\delta=0.001$ 
proved to be insufficient. 
Unfortunately, generating $p(N_f)$ distributions with this precision 
is very slow and a better, time delayed or targeted mechanism would be 
needed to see possible scaling of cascade sizes .

%%%%%%%%%%%%%%%%%%%%%%%%%%%%%%%%%%%%%%%%%%%%%%%%%%%%%%%%%%%%%%%%%%%%%%%%
%%%%%%%%%%%%%%%%%%%%%%%%%%%%%%%%%%%%%%%%%%%%%%%%%%%%%%%%%%%%%%%%%%%%%%%%

%%%%%%%%%%%%%%%%%%%%%%%%%%%%%%%%%%%%%%%%%%%%%%%%%%%%%%%%%%%%%%%%%%%%%%%%
\subsection{Summary of simulations}
%%%%%%%%%%%%%%%%%%%%%%%%%%%%%%%%%%%%%%%%%%%%%%%%%%%%%%%%%%%%%%%%%%%%%%%%

In this section we have shown results of extended dynamical simulations of 
the threshold synchronization, modeling power-grids on different topologies. 
We have found numerical evidences that line failure distributions can exhibit 
PL tails, in agreement with real statistics, by applying the second order 
threshold Kuramoto model on heterogeneous networks.
Although the second order Kuramoto model itself exhibits a discontinuous 
transition from chaotic to partially synchronized state, by increasing 
the oscillator couplings the desynchronization cascade size distributions 
of the threshold version show dynamical critical like behavior at and
below the transition, similar to what was obtained by SOC DC models earlier. 
This can happen if the heterogeneity in the system is moderately strong. 
For low heterogeneity, as in case of the square lattice, we have not found 
signatures of scale-free tails. For too strong heterogeneity, as in case of 
the HU-HV model with real inertia, the the level of synchronization remained 
very low. This could be compensated by equalizing the inertia terms or by 
a feedback mechanism. We have also shown that the application of 
exponentially distributed self-frequencies do not alter the exponents 
of PL tails, but of course decrease the synchronization order parameter. 
Thus they pose a moderate risk on the stability of power-grids.

%%%%%%%%%%%%%%%%%%%%%%%%%%%%%%%%%%%%%%%%%%%%%%%%%%%%%%%%%%%%%%%%%%%%%%%%
\section{Conclusions}
%%%%%%%%%%%%%%%%%%%%%%%%%%%%%%%%%%%%%%%%%%%%%%%%%%%%%%%%%%%%%%%%%%%%%%%%

Power-grids are becoming more and more heterogeneous as renewable 
(solar, wind, ... etc.) small suppliers are connected. 
Therefore, the danger of failures caused by desynchronization is of a 
great concern. Failure data of large power-grids have shown blackout 
size distributions with power-law (PL) tails. 
Previous simulations could explain this using power threshold cascade 
models, assuming self-organized criticality. 
In these DC models, the power redistribution, following a line or node cut, 
is described by a fixed amount of load. We have studied the stability 
of phase and frequency synchronized steady-states of realistic, 
Hungarian and US high voltage power grids using dynamical simulations of the
swing-equations, which describe the real power redistribution in AC 
electric networks. Earlier we have shown that heterogeneity can 
generate power-law desynchronization duration distributions without 
the assumption of criticality~\cite{POWcikk}.

Now we obtained roughly universal PL failure tails, without fine tuning 
to a critical point: i.e. at different thresholds (T), global couplings (K), 
and self-frequency distributions, for the 4941 node US and the 418 node 
HU-HV networks. The fitted exponents agree with those of the HU 
failure time data and other world-wide measurements. While the
synchronization values dropped, both the US and the HU grid cascade
size distributions seem to be insensitive to such stronger fluctuations.

We emphasize, that we don't rule out a SOC mechanism, which tunes 
the network into the neighborhood of the synchronization transition point,
as the consequence of power supply/demand competition, but show that
this parameter region is extended, due to heterogeneity and load 
dependent PL exponents may arise. The lack of PL-s in case of the
homogeneous 2D square lattice shows that heterogeneity must be taken into account,
simple homogeneous models cannot describe scale-free behavior of outages.  

We also found that too strong heterogeneity of inertia destabilizes 
the power-grid and reliable synchronization cannot be sustained without
feedback. Applying simple zero lag feedback were insufficient in our model,
possibly a time-delayed feedback control would be necessary as suggested 
in~\cite{Olmi-Sch-19}, which should be the target of further research.
This feedback is supposed to represent the frequency response of 
generators and loads. In case of generators, units providing primary 
reserve (or Frequency Containment Reserve) provide a practically 
immediate response based on the steepness [MW/Hz] of their open-loop 
control characteristic. Similarly, behavior of loads during frequency 
disturbances can be described by their respective correlation 
factor [MW/Hz], however their response is usually slightly delayed.
Still, without the this feedback our model is capable to describe
short time scales, which can be interesting for high variability 
systems with rapid changes, coming from large fluctuations of
renewable resources. 

Our future work will focus on further extensions of the presented model. 
To describe stabilization of synchronization a retarded model will be 
implemented, which has larger inertial feedback, thus compensating for 
the decrease of inertia due to renewable generation. 
Using the methods of complex network and hybrid tools a better insight can 
be gained into robustness and vulnerability issues of power systems. 
Such multi-level network analysis has been proven useful previously, 
as coupling level of different networked infrastructures may increase 
and decrease stability, depending on the actual level. 
Literature is yet to provide a validation of European power system 
failures in the presence of large share of distributed generation. 
It was also shown that tools, specifically designed for power system 
analysis outperform the methods that are solely built on topological 
connections. This gap between the two approaches is to be examined 
in detail by the authors, taking into consideration realistic network 
topologies and power flows, extreme failure statistics and the 
theory of self-critical systems.

The data-sets generated during and/or analyzed during the current study are
available from the corresponding author on reasonable request.

\acknowledgments{

We thank Benjamin Sch\"afer, Jeffrey Kelling and Benjamin Carreras  
for the useful comments.}

\funding{ Support from the MTA-EK special grant and the
Hungarian National Research, Development and Innovation
Office NKFIH (K109577) is acknowledged.
The VEKOP-2.3.2-16-2016-00011 grant is supported by the European Structural
and Investment Funds jointly financed by the European Commission and the
Hungarian Government.
Most of the numerical work was done on NIIF supercomputers of Hungary.
}

\authorcontributions{
G. \'O. wrote, ran and analyzed the Kuramoto model programs, 
and performed graph topology analysis. 
B. Hartmann invented the generalized Kuramoto equation to 
describe composite nodes, with more realistic parameters, 
collected network and failure data from the Hungarian electric 
company MAVIR. G. \'O. and B. Hartmann wrote the text.
B. Hartmann prepared Tables 1,2 and Figures 1,2.
G. \'O. prepared Figures 3-13.
All authors reviewed the manuscript.
}

\conflictsofinterest{ The authors declare no conflict of interest.}

\reftitle{References}

%%%%%%%%%%%%%%%%%%%%%%%%%%%%%%%%%%%%%%%%%%%%%%%%%%%%%%%%%%%%%%%%%%%%%%%%
\bibliography{bib}
%%%%%%%%%%%%%%%%%%%%%%%%%%%%%%%%%%%%%%%%%%%%%%%%%%%%%%%%%%%%%%%%%%%%%%%%

\end{document}